\documentclass[aps,prl,reprint,superscriptaddress]{revtex4-2}
\usepackage[T1]{fontenc}
\usepackage{bm}

\usepackage[table,xcdraw]{xcolor}
\usepackage{graphicx}
\usepackage{graphics}
\usepackage{amsmath,amssymb,amstext}
\usepackage{amsfonts}

\usepackage{hyperref}
\hypersetup{
 colorlinks,%
 citecolor=blue,%
 linkcolor=blue,%
 urlcolor=blue
}
\usepackage{tikz}\usetikzlibrary{petri}

\begin{document}

\newcommand{\up}{ \mathbb{\uparrow}}
\newcommand{\down}{ \mathbb{\downarrow}}
\newcommand{\Gmat}{ \mathbb{G}}
\newcommand{\deltaG}{ \Delta \mathbb{G}}
\newcommand{\bsigma}{\boldsymbol{\sigma}}
\newcommand{\Beff}{\textbf{B}_{\textbf{eff}}}

\newcommand{\ket}[1]{\vert#1\rangle }
\newcommand{\bra}[1]{\langle#1\vert}


\title{Impact of electrostatic crosstalk on spin qubits in dense CMOS quantum dot arrays}

\author{Jesus D. Cifuentes} 
\email{j.cifuentes_pardo@unsw.edu.au}
\affiliation{School of Electrical Engineering and Telecommunications, University of New South Wales, Sydney 2052, Australia}

\author{Tuomo Tanttu}
\affiliation{School of Electrical Engineering and Telecommunications, University of New South Wales, Sydney 2052, Australia}
\affiliation{Diraq, University of New South Wales, Sydney 2052, Australia}

\author{Paul Steinacker}
\affiliation{School of Electrical Engineering and Telecommunications, University of New South Wales, Sydney 2052, Australia}

\author{Santiago Serrano}
\affiliation{School of Electrical Engineering and Telecommunications, University of New South Wales, Sydney 2052, Australia}

\author{Ingvild Hansen}
\affiliation{School of Electrical Engineering and Telecommunications, University of New South Wales, Sydney 2052, Australia}

\author{James P. Slack-Smith}
\affiliation{School of Electrical Engineering and Telecommunications, University of New South Wales, Sydney 2052, Australia}

\author{Will Gilbert}
\affiliation{School of Electrical Engineering and Telecommunications, University of New South Wales, Sydney 2052, Australia}
\affiliation{Diraq, University of New South Wales, Sydney 2052, Australia}

\author{Jonathan Y. Huang}
\affiliation{School of Electrical Engineering and Telecommunications, University of New South Wales, Sydney 2052, Australia}

\author{Ensar Vahapoglu}
\affiliation{School of Electrical Engineering and Telecommunications, University of New South Wales, Sydney 2052, Australia}
\affiliation{Diraq, University of New South Wales, Sydney 2052, Australia}

\author{Ross C. C. Leon}
\affiliation{School of Electrical Engineering and Telecommunications, University of New South Wales, Sydney 2052, Australia}

\author{Nard Dumoulin Stuyck}
\affiliation{School of Electrical Engineering and Telecommunications, University of New South Wales, Sydney 2052, Australia}
\affiliation{Diraq, University of New South Wales, Sydney 2052, Australia}

\author{Kohei Itoh} 
\affiliation{School of Fundamental Science and Technology, Keio University, Yokohama, Japan}

\author{Nikolay Abrosimov}
\affiliation{Leibniz-Institut f\"{u}r Kristallz\"{u}chtung ,12489, Berlin, Germany}

\author{Hans-Joachim Pohl}
\affiliation{VITCON Projectconsult GmbH, 07745, Jena ,Germany}

\author{Michael Thewalt}
\affiliation{Department of Physics, Simon Fraser University, V5A 1S6, British Columbia, Canada}

\author{Arne Laucht}
\affiliation{School of Electrical Engineering and Telecommunications, University of New South Wales, Sydney 2052, Australia}
\affiliation{Diraq, University of New South Wales, Sydney 2052, Australia}

\author{Chih Hwan Yang}
\affiliation{School of Electrical Engineering and Telecommunications, University of New South Wales, Sydney 2052, Australia}
\affiliation{Diraq, University of New South Wales, Sydney 2052, Australia}

\author{Christopher C. Escott}
\affiliation{School of Electrical Engineering and Telecommunications, University of New South Wales, Sydney 2052, Australia}
\affiliation{Diraq, University of New South Wales, Sydney 2052, Australia}

\author{Fay E. Hudson}
\affiliation{School of Electrical Engineering and Telecommunications, University of New South Wales, Sydney 2052, Australia}
\affiliation{Diraq, University of New South Wales, Sydney 2052, Australia}

\author{Wee Han Lim}
\affiliation{School of Electrical Engineering and Telecommunications, University of New South Wales, Sydney 2052, Australia}
\affiliation{Diraq, University of New South Wales, Sydney 2052, Australia}

\author{Rajib Rahman}
\affiliation {School of Physics, University of New South Wales, Sydney 2052, Australia}

\author{Andrew S. Dzurak} 
\affiliation{School of Electrical Engineering and Telecommunications, University of New South Wales, Sydney 2052, Australia}
\affiliation{Diraq, University of New South Wales, Sydney 2052, Australia}

\author {Andre Saraiva}
\email{a.saraiva@unsw.edu.au}
\affiliation{School of Electrical Engineering and Telecommunications, University of New South Wales, Sydney 2052, Australia}
\affiliation{Diraq, University of New South Wales, Sydney 2052, Australia}



\date{ \today }
\begin{abstract}



Quantum processors based on integrated nanoscale silicon spin qubits are a promising platform for highly scalable quantum computation. Current CMOS spin qubit processors consist of dense gate arrays to define the quantum dots, making them susceptible to crosstalk from capacitive coupling between a dot and its neighbouring gates. Small but sizeable spin-orbit interactions can transfer this electrostatic crosstalk to the spin g-factors, creating a dependence of the Larmor frequency on the electric field created by gate electrodes positioned even tens of nanometers apart. By studying the Stark shift from tens of spin qubits measured in nine different CMOS devices, we developed a theoretical frawework that explains how electric fields couple to the spin of the electrons in increasingly complex arrays, including those electric fluctuations that limit qubit dephasing times $T_2^*$. The results will aid in the design of robust strategies to scale CMOS quantum technology.

\end{abstract} 

\maketitle

The vast historical investment in the development of transistors and integrated circuits could also play a pivotal role in the quantum era. Advances in nanotechnology based on silicon and its oxide recently enabled high-fidelity spin qubits~\cite{tanttu_stability_2023}, that can be operated at temperatures above 1K~\cite{huang2023highfidelity,Yang2020}. Similar qubit platforms built in silicon and germanium hetero-structures are also showing significant progress~\cite{noiri_fast_2022,mills_two-qubit_2022,camenzind_hole_2022,xue_quantum_2022}. In a constantly growing semiconductor quantum industry, the number of qubits per chip is increasing~\cite{philips_universal_2022, hendrickx_four-qubit_2021}. Million-dollar equipment, foundry-level fabrication techniques, and extensive knowledge accumulated through the last decades are now being applied to the development of quantum processors~\cite{zwerver_qubits_2022,elsayed_low_2022,neyens2023probing,Sabbagh2019}. As practical quantum advantage will likely demand the integration of millions of qubits for the execution of fault-tolerant error correcting codes~\cite{gidney_how_2021, beverland_assessing_2022}, this technology presents a promising pathway for the consolidation of quantum computation.






 \begin{figure}[t]
 \centering
 \includegraphics[width=\linewidth]{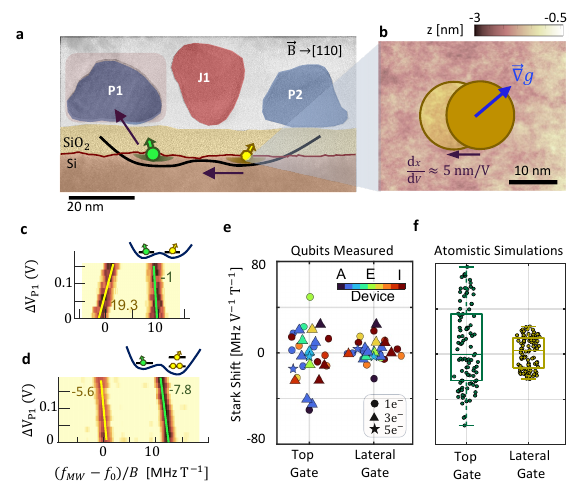}
 \caption{ \textbf{Models and measurements of spin-orbit crosstalk:}   \textbf{a}, Colored transmission electron microscopy (TEM) image of a MOS spin qubit device similar to the ones used in these experiments. Quantum dots are formed at the Si/SiO$_2$ interface below gates P1 and P2. A potential bias applied to gate P1 drags the qubit below P2. \textbf{b}, The change in surface profile generates a Stark shift ${\frac{dg_2}{dV_{P1}}= \frac{dx}{dV} \cdot \vec{\nabla}g}$ (see equation~\eqref{eq:crosstalk}). \textbf{c-d}, Qubit spectroscopy measurement \cite{Tanttu2019}, showing the Stark shift tunability of the Larmor frequency of qubits measured in Device I at double quantum dots with electron numbers: \textbf{c}, (1, 1)  and  \textbf{d} (1, 3) . \textbf{e}, Stark shifts of 30 qubits measured in 9 devices from top and lateral gates. (See Table~\ref{tab:StarkShifts}). \textbf{g}, We compare them with atomistic simulations of MOS quantum dots in a rough SiO$_2$ interface. $\vec{B}$ points to the $[110]$ lattice orientation in these measurements.  }
 \label{fig1}
\end{figure}


Silicon stands out as an excellent qubit platform. Its relatively small spin-orbit coupling protects the spin of the confined electrons from unavoidable electric fluctuations in the device~\cite{Saraiva2022, Tanttu2019} and contributes to a small variability of the Larmor frequency of each qubit ($<$1$\%$), which relaxes the requirements for the scalability of the technology~\cite{cifuentes_bounds_2023}.




Current control protocols in the field rely on these intrinsic variations between qubit frequencies for individual addressability. To control a specific qubit, for instance, a variable microwave field aligned with its unique Larmor resonance frequency is pulsed to induce gate rotations. These approaches have been useful for devices with small qubit numbers. They are, however, impractical for scaled processors as the frequency range of the qubits is limited, and a large number of qubits would be susceptible to frequency crowding~\cite{cifuentes_bounds_2023}.





A more scalable alternative proposes that multiple qubits can be driven in the global control field generated, for instance, by a dielectric resonator~\cite{hansen_pulse_2021,Vahapoglu2020}. For this to work, all qubits must have approximately the same frequency, aligned with the resonator frequency. In this case, individual addressing must be done spatially, by biasing a local gate that is close enough to shift the Larmor frequency of the target qubit~\cite{hansen_pulse_2021,seedhouse_quantum_2021} (see Fig.~\ref{fig1}\textbf{a-d}).

Despite the small spin-orbit effect in silicon conduction band electrons, their frequencies still present a measurable electrical tunability. This effect is a form of Stark shift, and it is a key part of the spin qubit technology as it allows us to address qubits through this electrostatic control~\cite{hansen_implementation_2022}. While all qubits need to have some Stark shift for this protocol to work, it is important to keep in mind that a large electric tunability exposes qubits to electric noise coming from two-level-fluctuators (TLFs) in the oxide stack~\cite{elsayed_low_2022, kane_silicon-based_1998}, potentially affecting their quality~\cite{Tanttu2019}.

Finding a good balance between these two situations requires an in-depth understanding of the microscopic origins of this Stark effect. So far, qubit Stark shifts have shown significant variations, even between qubits in the same device just a few nanometers apart~\cite{Ferdous2018}. The same dot may also yield different frequencies for qubits implemented with the spins of electrons in the outer shell of a multielectron dot configuration~\cite{Leon2020}  (See Fig.~\ref{fig1}\textbf{c-d}). In a recent paper, we attributed this variability to the presence of disorder in the system, here dominated by the roughness of the Si/SiO$_2$ interface~\cite{cifuentes_bounds_2023}. In this paper, we take a more detailed look into the microscopic origin of this electric tunability, the emergence of crosstalk terms from nearby gates and its impact on the susceptibility of qubits to charge noise.


We have now accumulated knowledge from a significant number of devices to explain the variations in the intrinsic Stark shift of electron spin qubits. This paper contains experimental data of the resonance frequency of 30 qubits measured in 9 MOS devices across 4 years of experiments. This includes data from qubits formed at the outer shell of 1, 3 and 5 electron quantum dots~\cite{Leon2020,Veldhorst2015}. Their spin-orbit coupling has a significant variability, with typical numbers for qubit Stark shifts ranging between -80 to 80~MHz~T$^{-1}$ V$^{-1}$ (see Fig.~\ref{fig1}\textbf{e}). This data was taken from pairs of qubits in double quantum dots formed in three dot and four dot devices (See Table \ref{tab:StarkShifts}). 

Significant Stark shifts can be also induced by neighbouring gates (Fig.~\ref{fig1}.\textbf{a-e}). This Stark shift-based electrostatic crosstalk is similar to other standard effects induced by gate proximity in quantum dot devices, such as cross-capacitive couplings~\cite{lawrie_spin_2020}, or the lever arms in the dot energy levels~\cite{hollmann_large_2020}, that can be driven by voltage biases applied to the nearby gates. The difference between these effects and spin-orbit crosstalk is that the last is more variable due to the strong dependence of the qubit g-factors on the surface profile. The size and the sign of the Stark shifts is dominated by the relative displacement of the dots across the rough interface (see Fig.~\ref{fig1}.\textbf{b}). This idea is supported by atomistic tight-binding quantum dot simulations with interface roughness profiles(see Fig.~\ref{fig1}.\textbf{f})~\cite{cifuentes_bounds_2023,Klimeck2007,Naumov2008}. The methods for modeling this system are described in the following section.

\subsection{Model of spin-orbit tunability}

To understand better this problem we developed a simple model for g-factor tunability based on the methods from a previous paper~\cite{cifuentes_bounds_2023}. We performed electrostatic simulations of realistic 3D devices in COMSOL Multiphysics and we fitted the quantum dot region to a harmonic model with a vertical electric field
\begin{equation} \label{eq2}
 V(x,y,z) = c_x(x-x_c)^2 + c_y(y-y_c)^2 + zE_z, 
\end{equation}
\noindent where $(x_c, y_c)$ is the centre of the parabolic potential, $E_z$ is the electric field in the z-axis ($\sim$ 20 mV nm$^{-1}$) and $c_x, c_y$ are the lateral curvatures ($\sim$ 0.3 mV nm$^{-2}$). This is a good approximation in these devices near the potential minimum of quantum dots that are far apart enough from each other to neglect exchange interactions. Sweeps in potential gate biases simulated in Comsol lead to linear changes in the fitted dot parameters that can be characterized by $\frac{dx_c}{dV}$, $\frac{dy_c}{dV}$ and $\frac{dE_z}{dV}$. The shifts in the $x_c$ and $y_c$ can be converted to in-plane electric fields in the harmonic approximation as $\frac{dx_c}{dV} = \frac{1}{2c_x} \frac{dE_x}{dV} $. The best fit for the experimental data, in Fig.~\ref{fig1}.\textbf{c} is given by taking $\frac{dE_z}{dV} = 11$~mV/nm and $\frac{dx_c}{dV} = 5$~nm/V, which are similar to the values estimated by electrostatic simulations of this device (see Table~\ref{tab:PotentialSweeps} in the Supporting Information).

In this approach, we study the simplified case of the g-factor Stark shift and later we tackle the more general Stark shift structure of the g-tensor. We can estimate that the Stark shift of gate $j$ on qubit $n$ is to first order

\begin{equation}
\label{eq:crosstalk}
 \frac{\partial g_{n}}{\partial V_{j}} = \frac{\partial \vec{E}_n}{\partial V_{j} } \cdot \vec{\nabla}_{\vec{E}_n} \left( g_n \right),
\end{equation}
or,

\begin{align}
 \frac{\partial g_{n}}{\partial V_{j}} = \frac{\partial E_{n,x}}{\partial V_{j}}\frac{\partial g_{n}}{\partial E_{n,x}}+\frac{\partial E_{n,y}}{\partial V_{j}}\frac{\partial g_{n}}{\partial E_{n,y}}+\frac{\partial E_{n,z}}{\partial V_{j}}\frac{\partial g_{n}}{\partial E_{n,z}}, 
\end{align}

\noindent which leads to a matrix of Stark shifts $\frac{\partial g_{n}}{\partial V_{j}}$. The equation divides the Stark shift of these qubits into two terms.

The first depicts how effective a gate $j$ is in generating electric fields in quantum dot $n$, $\frac{\partial \vec{E}_n}{\partial V_j}$ (purple arrows in Fig.~\ref{fig1}\textbf{d}). This is entirely determined by the device layout. Any changes to this term (e.g. the thickness of the oxide, the size of the gates, array organization), would lead to modifications on the controllability of the dot parameters. 
We estimate this part with electrostatic simulations in COMSOL Multiphysics (see Table~\ref{tab:PotentialSweeps}). 


The second part accounts for the changes in the g-factor of dot $n$ when it is pushed towards a certain direction ($\vec{\nabla }g_n$) (blue vectors Fig.~\ref{fig1}\textbf{d}). This accounts for the random behavior of the Stark shifts when the dots are displaced in a disordered environment. We can estimate its variability with atomistic tight binding simulations of quantum dots under a rough Si/SiO$_2$ interface.

The main consequence of equation \eqref{eq:crosstalk} is that the vectors $\frac{\partial \vec{E}}{\partial V_{j} }$ and $\vec{\nabla }g$ must be aligned for the Stark shift to be positive. If they are anti-aligned it would be negative, and if they are orthogonal, the qubit Stark shifts would vanish. Also, Stark shifts would no be observed in the random cases where $\vert \vec{\nabla}g \vert$ is too small. All these situations are observed with different probabilities in both experimental and simulation data (Fig.~\ref{fig1}\textbf{e-f}).

\begin{figure*}[hbt]
 \centering
 \includegraphics[width=\linewidth]{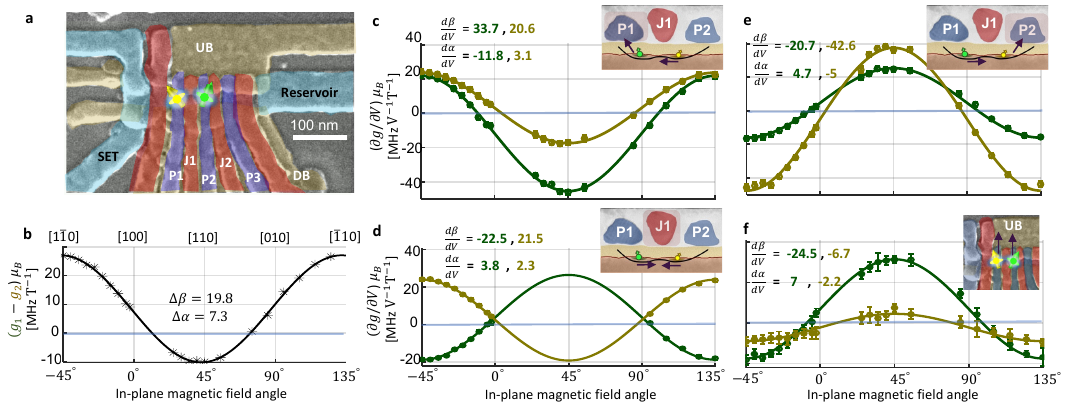}
 \caption{\textbf{Stark shift tunability of the two qubits in device A from a variety of gates:} \textbf{a}, SEM of a device with the same design as device A. Two qubits were formed below gates P1 and P2. \textbf{b}, Dependence of the difference between the g-factors of the two qubits versus the in-plane magnetic field direction. \textbf{c-f}, Dependence versus in-plane magnetic field of the Stark shift on gates: \textbf{c}, P1 ; \textbf{d}, J1 ; \textbf{e}, P2; \textbf{f}, UB. The numbers in the figures show the fits for Dresselhaus ($\beta$) and Rashba ($\alpha$) spin-orbit-coupling, as well as for the voltage dependencies $\partial \alpha /\partial V_{j}$ , $\partial \beta /\partial V_{j}$ for each gate $g\in$ \{P1, P2, J1, UB\}. The units of these insets are the same as the variable measured in each plot. }
 \label{fig2}
\end{figure*}

\begin{figure}[b]
 \centering
 \includegraphics[width=\linewidth]{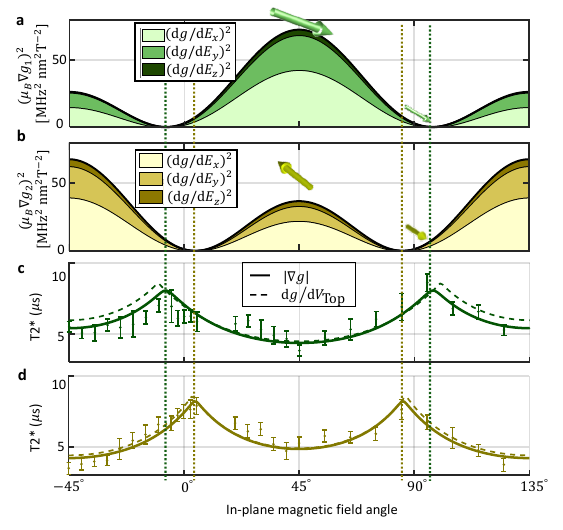}
 \caption{ \textbf{Stark shifts vs T$_2^*$}: \textbf{a-b}, Decomposition of g-gradients $\Vec{\nabla} g_{1(2)}$ of both qubits in FIG.\ref{fig2} vs in-plane magnetic field direction. \textbf{c-d}, T$_2^*$ measurements of these qubits. The fits are performed using two models: Dashed lines assume that the electric noise comes only from the top gate. Solid lines assume that electric noise can come from any orientation with equal probability. }
 \label{fig3}
\end{figure}



\subsection{Spin-orbit tunability \textbf{versus} magnetic field angles}


The microscopic origin of the spin-orbit coupling is revealed when the magnetic field direction is swept~\cite{Tanttu2019}. The asymmetry caused by the Si/SiO$_2$ interface against which the quantum dots are formed, combined with the crystal structure of the silicon lattice, lead to a g-factor dependence on the magnetic field direction~\cite{Ruskov2018,Ferdous2018}. This dependence is described in the $\mathbb{G}$-matrix formalism, where the Zeeman Hamiltonian is written as 
\begin{equation}
\label{eq:Hzeeman}
H_{\rm Zeeman} = \frac{\mu_B}{2} \vec{\sigma}^T \mathbb{G} \vec{B}. 
\end{equation}
We have previously investigated atomistically the form of this $\mathbb{G}$-matrix for silicon quantum dots~\cite{cifuentes_bounds_2023}, and obtained 
\begin{equation}
\label{eq:Gmat1}
		\mathbb{G} = 
		 \begin{bmatrix}
			g_0+ 4\alpha & 4\beta & g_{13}\\
			4\beta & g_0 + 4\alpha & g_{23} \\
			0 & 0 & g_{33}
			\end{bmatrix}, 
\end{equation}
\noindent where the basis vectors in which this $\mathbb{G}$-matrix is written are aligned with the cubic lattice orientations $\{[100], [010], [001]\}]$. Typical values obtained from atomistic simulation are $g_0 \approx 1.9937$, $\alpha' \sim -10^{-3}$, $\beta' \sim \pm 10^{-2}$ and $g_{13} \sim \pm 10^{-3}, g_{23} \pm \sim 10^{-3} $ and $ g_{33} \sim 2.00192 - \mathcal{O}(10^{-4}) $. For any magnetic field direction $\hat{r}$, the g-factor is ${g(\hat{r}) = \Vert \mathbb{G} \hat{r}\Vert}$.

For an in-plane magnetic field forming an angle $\varphi$ with the $[100]$ lattice orientation, and given that $\alpha$ and $ \beta \ll g_0$, ${g\left(\varphi\right) \approx g_0+\alpha+\beta\sin{\left(2\varphi\right)} }$~\cite{cifuentes_bounds_2023}. These two parameters $\alpha$ and $\beta$ are usually associated with Rashba and Dresselhaus spin-orbit coupling. Interface roughness affects both of them through different mechanisms, implying that $\alpha$ and $\beta$ are effectively independent random variables. The variability range of $\beta$ ($\sim 50$ MHz/T) is one order of magnitude larger than $\alpha$ ($\sim 5$ MHz/T). However, observing these two effects in a single experiment in device A (see Fig.~\ref{fig2}\textbf{a-b}), we notice that Rashba still has an important contribution. Because of this term, the angle at which any pair of qubit frequencies becomes identical is slightly deviated from the [100] lattice orientation. In this case, it is detuned by a difference of $10$MHz T$^-1$ caused by the variability in Rashba spin-obit coupling. 



Similar dependencies are observed in the Stark shifts 
\begin{equation}
\label{eq:StarkInplane}
   \frac{\partial g_n \left(\varphi\right)}{\partial V_{j}}=\frac{\partial \alpha_n}{\partial V_{j}}+\frac{\partial \beta_n}{\partial V_{j}}\sin{\left(2\varphi\right)},
\end{equation}
\noindent with the sole distinction that each gate $j$ should have a different impact on $\alpha$ and $\beta$ (see Fig.\ref{fig2}\textbf{c-f}). We fitted $\frac{\partial \alpha_{n}}{\partial V_{j}}$ and $\frac{\partial \beta_{n}}{dV_{j}}$ for all gates $j$ and for both qubits $n= 1,2$ and wrote the result on the inset of each figure. Equation~\eqref{eq:crosstalk} also applies for each spin-orbit parameter
\begin{equation}
   \label{eq:alpha}
    \frac{\partial \alpha_{n}}{\partial V_{j}} = \frac{\partial \vec{E}_{n}}{\partial V_{j} } \cdot \vec{\nabla}_{\vec{E}} \left( \alpha_n \right) \ \text{ and } \ \frac{d \beta_{n}}{dV_{j}}= \frac{\partial \vec{E}_{n}}{\partial V_{j} } \cdot \vec{\nabla}_{\vec{E}} \left( \beta_n \right),
\end{equation}
which will allow us to estimate the changes in g-factor gradient of the two qubits measured in the experiment across a varying in-plane field ${ \vec{\nabla} g(\varphi)=\vec{\nabla}\alpha+\vec{\nabla} \beta \sin 2 \varphi }$.


We start by understanding the action of the gates on both quantum dots $\frac{\partial \vec{E}}{\partial V_{j} } $. Side plunger gates ($\frac{\partial g_1}{\partial V_{P2}}$ and $\frac{\partial g_2}{\partial V_{P1}}$ ) and J-gates are the simplest case as they do not impact the electric confinement in the z-axis (Table~\ref{tab:PotentialSweeps}), so their contribution can be approximated to a lateral displacement of the dot center in the x-axis towards the other qubit (see inset figures in Fig.~\ref{fig2}\textbf{c-f}). Because of this, Stark shifts performed by side gates and J-gates usually have the same sign. In this particular experiment $\frac{\partial g_1}{\partial V_{P2}} \sim \frac{\partial g_1}{\partial V_{J1}}$ and $\frac{\partial g_2}{\partial V_{P1}} \sim \frac{\partial g_2}{\partial V_{J1}}$ as seen in Fig.~\ref{fig2}\textbf{c-e}. This feature allows us to effectively characterize which frequency corresponds to each dot, even under the presence of crosstalk.


Other gates in the device have more convoluted impacts on the quantum dots. The plunger gates on top of the qubits ($\frac{\partial g_1}{\partial V_{P1}}$ and $\frac{\partial g_2}{\partial V_{P2}}$) combine a lateral movement in the opposite x direction with a sizeable increase in the out-of-plane electric confinement $E_z$ . Upper barrier (UB) gates do something similar, but this time the main displacement is towards the $y$-axis. We can quantify this information in the following matrix, which is based on the results from Comsol simulations, and the parameters that provide the best fit to Fig.~\ref{fig1}.\textbf{c}
\begin{equation}
\label{eq:DE}
\mathbf{D}_V ( \vec{E} ) :=\left[\begin{array}{ccc}
\frac{\partial E_{x}}{\partial V_{\text{Top}}} & \frac{\partial E_{y}}{\partial V_{\text{Top}}} & \frac{\partial E_{z}}{dV_{\text{Top}}}\\
\frac{\partial E_{x}}{\partial V_{\text{side}}} & \frac{\partial E_{y}}{\partial V_{\text{side}}} & \frac{\partial E_{z}}{dV_{\text{side}}}\\
\frac{\partial E_{x}}{\partial V_{\text{UB}}} & \frac{\partial E_{y}}{\partial V_{\text{UB}}} & \frac{\partial E_{z}}{dV_{\text{UB}}}
\end{array}\right]=\left[\begin{array}{ccc}
-3.5 & 0 & 11\\
4.5 & 0 & 2\\
0 & 4 & -5
\end{array}\right], 
\end{equation}
\noindent with units on mV~nm$^{-1}$~V$^{-1}$. Here we do not include a row for $\partial g/\partial J$ as its impact is similar to lateral gates. By inverting ${\mathbf{D}_V ( \vec{E} )}$, and multiplying it by the array of $\frac{\partial \alpha_{n}}{\partial V_{j}}$ and $\frac{\partial \beta_{n}}{\partial V_{j}}$ for each qubit $n={1,2}$, we obtain the g-factor gradients $\vec{\nabla}\alpha_n $ and $\vec{\nabla} \beta_n$ in equation \eqref{eq:alpha}.

The three-dimensional decomposition of $\vec{\nabla} g_1$ and $\vec{\nabla} g_2$ is plotted in Fig.~\ref{fig3}.\textbf{a-b}. The direction and magnitude of the g-factor gradient change with the in-plane angle. Vectors $\vec{\nabla} g(0^\circ)= \vec{\nabla}\alpha $ and $\vec{\nabla} g(45^\circ)= \vec{\nabla}\alpha +\vec{\nabla}\beta$ are independent variables, which assume values depending on the random surface roughness, as discussed previously.

In the analysis thus far, we assumed that the spin-orbit effects are only a small correction over the bulk $g_0$. It is also possible to make a more general description of these results in the $\mathbb{G}$-matrix formalism, considering that ${g(\hat{r}) = \Vert \mathbb{G} \hat{r}\Vert = \sqrt{ \hat{r}^T \mathbb{G}^T \mathbb{G} \hat{r}} } $. Then 
\begin{equation}
\frac{\partial g(\hat{r})}{\partial V_{j}}=\frac{\hat{r}^{T}\left(\frac{\partial \mathbb{G}^{T}}{\partial V_{j}}\mathbb{G}+\mathbb{G}^{T}\frac{\partial \mathbb{G}}{\partial V_{j}}\right)\hat{r}}{2g(\hat{r})}.
\end{equation}
\noindent This equation is more appropriate for systems with high spin-orbit coupling where the approximation $\alpha,\beta \ll g_0$ may not be valid.

 \begin{figure}[bt]
 \centering
 \includegraphics[width=\linewidth]{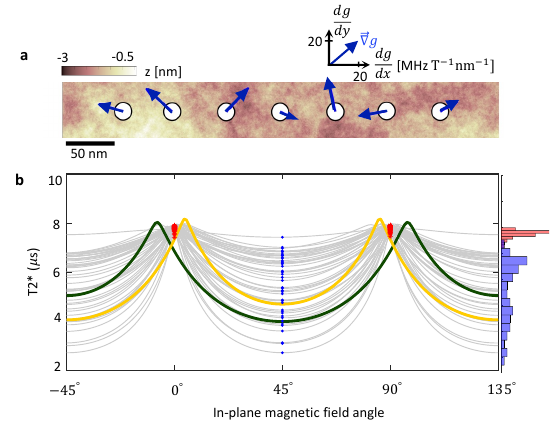}
 \caption{ \textbf{ Impact of variability of $\nabla g$ on the dephasing times T$_2^*$}. \textbf{a}, Seven spin qubits in a rough silicon oxide interface will have different $\vec{\nabla} g$. These vectors where obtained from atomistic simulations of quantum dots with their respective local surface profile. (B$\rightarrow$[110]) \textbf{b}, Expected variability of the dephasing times T$_2^*$ of MOS spin qubits. The histograms colored with blue and red are associated to the distribution of T$_2^*$  when the magnetic field points to $45^\circ$ ([110]) and to $0^\circ$ ([100]). The green and yellow traces correspond to the two qubits measured in Fig.~\ref{fig3}.}
 \label{fig4}
\end{figure}

\section{Qubit dephasing times}

The presence of electrical noise, including the inextricable $1/f$ noise caused by the oxide stack, causes the $T_2^*$ of these two qubits to be related to the norm of the gradient $\vec{\nabla}g_n$ (see Fig.~\ref{fig3}.\textbf{c-d}). The noise that limits the $T_2^*$ times in electron spin qubits can be divided in two types as
\begin{equation}
   \frac{1}{T_2^{\star}} = \frac{1}{T_0} + \frac{1}{T_e}.
\end{equation}
$T_0$ stands for magnetic noise (nuclear spins, paramagnetic fluctuations, magnetic drift, etc) and $T_e$ accounts for the electric noise produced by two-level-fluctuators~\cite{Ferdous2018}. This second type of noise couples to the qubit through Stark shift, so it depends on the in-plane magnetic angle $T_e (\phi)$. 

Previous models have assumed that this electric noise is dominated by noise in the top gate, such that $\frac{1}{T_e} = \Delta V \frac{\partial g}{\partial V_{\rm Top}}$~\cite{Tanttu2019}. However, looking at Fig.~\ref{fig2} it is clear that noise coming from other directions can also affect the qubits. To represent this, we replaced the top gate Stark shift by 
\begin{equation}
   \frac{1}{T_e} = \Delta E \mu_B \Vert  \vec{\nabla} g \Vert B, 
\end{equation}

\noindent where $B$ is the magnetic field and $\Delta E$ is a fitting parameter representing the amplitude of the electric noise. For this experiment, we considered $B=$0.82T. In Fig.~\ref{fig3}.\textbf{c-d} we can see that this last model fits the T$_2^*$ data slightly better than the top gate model. The values extracted for the fits are $T_0$ = 7.72 $\mu$s for the magnetic de-phasing times and $\Delta E = 0.017$ mV/nm. These numbers are within reasonable limits~\cite{Tanttu2019}, and assumed to be the same for both qubits. 

The differences between their coherence times are attributed to the spin-orbit coupling variations between qubits. Notice that the angles of maximum dephasing times coincide with the Stark shift minima of each spin. 

In a processor with many qubits, a magnetic field that is optimal for most of the qubits should be selected. Each qubit will have a different g-factor gradient $\vec{\nabla}g$ according to its surface profile (FIG.~\ref{fig4}\textbf{a}). The $\vec{B}$ field orientation that will maximize the qubit coherence for most of the qubits is [100] as seen in FIG.~\ref{fig4}\textbf{b}. This is also the direction at which the $T_2^*$ variability is minimized. This plot is done assuming that $T_0$ = 7.72 and $\Delta E = 0.017$ mV/nm are equal for all qubits.

In reality, both variables can have variability. Nuclear spin fluctuations can be different for both qubits\cite{zhao_single-spin_2019}, and qubits in the same substrate but well spaced from each other could be exposed to different levels of noise. As rotating the magnetic field allows to control the magnitude of the Dresselhaus spin-orbit interactions, it is a good method to separate electric effects from other sources of noise. If this experiment was performed in a larger amount of well-characterized qubits in the same substrate, the observed variability trends could help us to understand the microscopic origin of this electric noise, which is of special importance for quantum computation~\cite{Shehata2023,choi2023interacting}.

\section*{Conclusions }

We have provided a detailed quantitative explanation of how electric fields couple to MOS spin qubits due to spin-orbit interactions. The main conclusion is that the g-factor of each qubit is different due to the presence of local sources of disorder, and the magnitude and sign of the Stark shifts depend on which gate and towards which direction the electron is pushed. This model explains the Stark shift of tens of qubits measured in our facilities.

At a general level, this implies that the spin-orbit tuning that each gate performs on each qubit is different and must be characterized for every device. This must be done only once, as the qubit g-factors are stable enough in time, and only affected by time-dependent disorder at a much smaller scale ($\approx$ 0.01\%)~\cite{zhao_single-spin_2019,chan_assessment_2018}. To enable individual addressability in global control protocols, it will be necessary to characterize the Stark shift that each gate $(g)$ performs in the the qubits ($n$) nearby $(\frac{dg_n}{dV_g})$. This will enable the implementation of a linear combination of (virtual) gates to control each qubit ~\cite{unseld_2d_2023, borsoi_shared_2022,hendrickx_four-qubit_2021}.


The framework described in this paper works for all devices that we have measured so far. It only fails in the particular case when the qubits are exposed to a state transition that could be induced either by a valley or orbital degeneracies~\cite{gilbert_-demand_2023,Bourdet2018}. At this point the spin-orbit coupling is completely dominated by orbital mixing and the contribution of interface disorder is unknown.

We also studied the impact of these Stark shift variations on the qubit coupling to electric noise in a varying magnetic field direction. We found excellent agreement between the individual Stark shift tunability the qubits measured in Device A, with their corresponding T$_2^*$ profile. The results and methods provided in here can help us to understand better the composition of disorder that affects spin qubits between electric and non-electric sources. Extended versions of this approach in larger arrays of dots with well-characterized qubits in the same subtract could also help to study the location of two-level-fluctuators in silicon MOS devices. 

Finally, our results reveal that operating with the magnetic field pointing at the [100] lattice orientation provides multiple advantages for global control~\cite{hansen_pulse_2021}. Lower g-factor variability~\cite{cifuentes_bounds_2023}, smaller electric noise coupling to spin qubits, with consequently higher dephasing times and with smaller variation between qubits. Despite the smaller spin-orbit coupling, Fig.\ref{fig2} shows that there still remains an acceptable Stark shift tunability of both qubits from multiple gates, which is important for the practical implementation of this approach~\cite{hansen_implementation_2022}. 


\section*{Data Availability}
The datasets generated and/or analysed during this
study are available from the corresponding authors on
reasonable request.

\section*{Code Availability}
The analysis codes that support the findings of the
study are available from the corresponding authors on
reasonable request.

\section*{Acknowledgments}
We thank A. Dickie for technical help. We acknowledge support from the Australian Research Council (FL190100167 and CE170100012), the US Army Research Office (W911NF-23-1-0092), and the NSW Node of the Australian National Fabrication Facility. The views and conclusions contained in this document are those of the authors and should not be interpreted
as representing the official policies, either expressed or
implied, of the Army Research Office or the US Govern-
ment. The US Government is authorized to reproduce
and distribute reprints for Government purposes notwith-
standing any copyright notation herein. J.Y.H., P.S,  S.S., I.H, J.P.S.-S. and J.D.C. acknowledge support from the Sydney Quantum Academy. This project was undertaken with the assistance of resources and services from the National Computational Infrastructure (NCI), which is supported by the Australian Government.

\bibliography{CrosstalkPaper}

\begin{thebibliography}{39}%
\makeatletter
\providecommand \@ifxundefined [1]{%
 \@ifx{#1\undefined}
}%
\providecommand \@ifnum [1]{%
 \ifnum #1\expandafter \@firstoftwo
 \else \expandafter \@secondoftwo
 \fi
}%
\providecommand \@ifx [1]{%
 \ifx #1\expandafter \@firstoftwo
 \else \expandafter \@secondoftwo
 \fi
}%
\providecommand \natexlab [1]{#1}%
\providecommand \enquote  [1]{``#1''}%
\providecommand \bibnamefont  [1]{#1}%
\providecommand \bibfnamefont [1]{#1}%
\providecommand \citenamefont [1]{#1}%
\providecommand \href@noop [0]{\@secondoftwo}%
\providecommand \href [0]{\begingroup \@sanitize@url \@href}%
\providecommand \@href[1]{\@@startlink{#1}\@@href}%
\providecommand \@@href[1]{\endgroup#1\@@endlink}%
\providecommand \@sanitize@url [0]{\catcode `\\12\catcode `\$12\catcode
  `\&12\catcode `\#12\catcode `\^12\catcode `\_12\catcode `\%12\relax}%
\providecommand \@@startlink[1]{}%
\providecommand \@@endlink[0]{}%
\providecommand \url  [0]{\begingroup\@sanitize@url \@url }%
\providecommand \@url [1]{\endgroup\@href {#1}{\urlprefix }}%
\providecommand \urlprefix  [0]{URL }%
\providecommand \Eprint [0]{\href }%
\providecommand \doibase [0]{https://doi.org/}%
\providecommand \selectlanguage [0]{\@gobble}%
\providecommand \bibinfo  [0]{\@secondoftwo}%
\providecommand \bibfield  [0]{\@secondoftwo}%
\providecommand \translation [1]{[#1]}%
\providecommand \BibitemOpen [0]{}%
\providecommand \bibitemStop [0]{}%
\providecommand \bibitemNoStop [0]{.\EOS\space}%
\providecommand \EOS [0]{\spacefactor3000\relax}%
\providecommand \BibitemShut  [1]{\csname bibitem#1\endcsname}%
\let\auto@bib@innerbib\@empty
\bibitem [{\citenamefont {Tanttu}\ \emph {et~al.}(2023)\citenamefont {Tanttu},
  \citenamefont {Lim}, \citenamefont {Huang}, \citenamefont {Stuyck},
  \citenamefont {Gilbert}, \citenamefont {Su}, \citenamefont {Feng},
  \citenamefont {Cifuentes}, \citenamefont {Seedhouse}, \citenamefont
  {Seritan}, \citenamefont {Ostrove}, \citenamefont {Rudinger}, \citenamefont
  {Leon}, \citenamefont {Huang}, \citenamefont {Escott}, \citenamefont {Itoh},
  \citenamefont {Abrosimov}, \citenamefont {Pohl}, \citenamefont {Thewalt},
  \citenamefont {Hudson}, \citenamefont {Blume-Kohout}, \citenamefont
  {Bartlett}, \citenamefont {Morello}, \citenamefont {Laucht}, \citenamefont
  {Yang}, \citenamefont {Saraiva},\ and\ \citenamefont
  {Dzurak}}]{tanttu_stability_2023}%
  \BibitemOpen
  \bibfield  {author} {\bibinfo {author} {\bibfnamefont {T.}~\bibnamefont
  {Tanttu}}, \bibinfo {author} {\bibfnamefont {W.~H.}\ \bibnamefont {Lim}},
  \bibinfo {author} {\bibfnamefont {J.~Y.}\ \bibnamefont {Huang}}, \bibinfo
  {author} {\bibfnamefont {N.~D.}\ \bibnamefont {Stuyck}}, \bibinfo {author}
  {\bibfnamefont {W.}~\bibnamefont {Gilbert}}, \bibinfo {author} {\bibfnamefont
  {R.~Y.}\ \bibnamefont {Su}}, \bibinfo {author} {\bibfnamefont
  {M.}~\bibnamefont {Feng}}, \bibinfo {author} {\bibfnamefont {J.~D.}\
  \bibnamefont {Cifuentes}}, \bibinfo {author} {\bibfnamefont {A.~E.}\
  \bibnamefont {Seedhouse}}, \bibinfo {author} {\bibfnamefont {S.~K.}\
  \bibnamefont {Seritan}}, \bibinfo {author} {\bibfnamefont {C.~I.}\
  \bibnamefont {Ostrove}}, \bibinfo {author} {\bibfnamefont {K.~M.}\
  \bibnamefont {Rudinger}}, \bibinfo {author} {\bibfnamefont {R.~C.~C.}\
  \bibnamefont {Leon}}, \bibinfo {author} {\bibfnamefont {W.}~\bibnamefont
  {Huang}}, \bibinfo {author} {\bibfnamefont {C.~C.}\ \bibnamefont {Escott}},
  \bibinfo {author} {\bibfnamefont {K.~M.}\ \bibnamefont {Itoh}}, \bibinfo
  {author} {\bibfnamefont {N.~V.}\ \bibnamefont {Abrosimov}}, \bibinfo {author}
  {\bibfnamefont {H.-J.}\ \bibnamefont {Pohl}}, \bibinfo {author}
  {\bibfnamefont {M.~L.~W.}\ \bibnamefont {Thewalt}}, \bibinfo {author}
  {\bibfnamefont {F.~E.}\ \bibnamefont {Hudson}}, \bibinfo {author}
  {\bibfnamefont {R.}~\bibnamefont {Blume-Kohout}}, \bibinfo {author}
  {\bibfnamefont {S.~D.}\ \bibnamefont {Bartlett}}, \bibinfo {author}
  {\bibfnamefont {A.}~\bibnamefont {Morello}}, \bibinfo {author} {\bibfnamefont
  {A.}~\bibnamefont {Laucht}}, \bibinfo {author} {\bibfnamefont {C.~H.}\
  \bibnamefont {Yang}}, \bibinfo {author} {\bibfnamefont {A.}~\bibnamefont
  {Saraiva}},\ and\ \bibinfo {author} {\bibfnamefont {A.~S.}\ \bibnamefont
  {Dzurak}},\ }\href {https://doi.org/10.48550/arXiv.2303.04090} {\bibinfo
  {title} {Stability of high-fidelity two-qubit operations in silicon}}
  (\bibinfo {year} {2023}),\ \bibinfo {note} {arXiv:2303.04090 [cond-mat,
  physics:quant-ph]}\BibitemShut {NoStop}%
\bibitem [{\citenamefont {Huang}\ \emph {et~al.}(2023)\citenamefont {Huang},
  \citenamefont {Su}, \citenamefont {Lim}, \citenamefont {Feng}, \citenamefont
  {van Straaten}, \citenamefont {Severin}, \citenamefont {Gilbert},
  \citenamefont {Stuyck}, \citenamefont {Tanttu}, \citenamefont {Serrano},
  \citenamefont {Cifuentes}, \citenamefont {Hansen}, \citenamefont {Seedhouse},
  \citenamefont {Vahapoglu}, \citenamefont {Abrosimov}, \citenamefont {Pohl},
  \citenamefont {Thewalt}, \citenamefont {Hudson}, \citenamefont {Escott},
  \citenamefont {Ares}, \citenamefont {Bartlett}, \citenamefont {Morello},
  \citenamefont {Saraiva}, \citenamefont {Laucht}, \citenamefont {Dzurak},\
  and\ \citenamefont {Yang}}]{huang2023highfidelity}%
  \BibitemOpen
  \bibfield  {author} {\bibinfo {author} {\bibfnamefont {J.~Y.}\ \bibnamefont
  {Huang}}, \bibinfo {author} {\bibfnamefont {R.~Y.}\ \bibnamefont {Su}},
  \bibinfo {author} {\bibfnamefont {W.~H.}\ \bibnamefont {Lim}}, \bibinfo
  {author} {\bibfnamefont {M.}~\bibnamefont {Feng}}, \bibinfo {author}
  {\bibfnamefont {B.}~\bibnamefont {van Straaten}}, \bibinfo {author}
  {\bibfnamefont {B.}~\bibnamefont {Severin}}, \bibinfo {author} {\bibfnamefont
  {W.}~\bibnamefont {Gilbert}}, \bibinfo {author} {\bibfnamefont {N.~D.}\
  \bibnamefont {Stuyck}}, \bibinfo {author} {\bibfnamefont {T.}~\bibnamefont
  {Tanttu}}, \bibinfo {author} {\bibfnamefont {S.}~\bibnamefont {Serrano}},
  \bibinfo {author} {\bibfnamefont {J.~D.}\ \bibnamefont {Cifuentes}}, \bibinfo
  {author} {\bibfnamefont {I.}~\bibnamefont {Hansen}}, \bibinfo {author}
  {\bibfnamefont {A.~E.}\ \bibnamefont {Seedhouse}}, \bibinfo {author}
  {\bibfnamefont {E.}~\bibnamefont {Vahapoglu}}, \bibinfo {author}
  {\bibfnamefont {N.~V.}\ \bibnamefont {Abrosimov}}, \bibinfo {author}
  {\bibfnamefont {H.-J.}\ \bibnamefont {Pohl}}, \bibinfo {author}
  {\bibfnamefont {M.~L.~W.}\ \bibnamefont {Thewalt}}, \bibinfo {author}
  {\bibfnamefont {F.~E.}\ \bibnamefont {Hudson}}, \bibinfo {author}
  {\bibfnamefont {C.~C.}\ \bibnamefont {Escott}}, \bibinfo {author}
  {\bibfnamefont {N.}~\bibnamefont {Ares}}, \bibinfo {author} {\bibfnamefont
  {S.~D.}\ \bibnamefont {Bartlett}}, \bibinfo {author} {\bibfnamefont
  {A.}~\bibnamefont {Morello}}, \bibinfo {author} {\bibfnamefont
  {A.}~\bibnamefont {Saraiva}}, \bibinfo {author} {\bibfnamefont
  {A.}~\bibnamefont {Laucht}}, \bibinfo {author} {\bibfnamefont {A.~S.}\
  \bibnamefont {Dzurak}},\ and\ \bibinfo {author} {\bibfnamefont {C.~H.}\
  \bibnamefont {Yang}},\ }\href@noop {} {\bibinfo {title} {High-fidelity
  operation and algorithmic initialisation of spin qubits above one kelvin}}
  (\bibinfo {year} {2023}),\ \Eprint {https://arxiv.org/abs/2308.02111}
  {arXiv:2308.02111 [quant-ph]} \BibitemShut {NoStop}%
\bibitem [{\citenamefont {Yang}\ \emph {et~al.}(2020)\citenamefont {Yang},
  \citenamefont {Leon}, \citenamefont {Hwang}, \citenamefont {Saraiva},
  \citenamefont {Tanttu}, \citenamefont {Huang}, \citenamefont {Lemyre},
  \citenamefont {Chan}, \citenamefont {Tan}, \citenamefont {Hudson},
  \citenamefont {Itoh}, \citenamefont {Morello}, \citenamefont
  {Pioro-Ladrière}, \citenamefont {Laucht},\ and\ \citenamefont
  {Dzurak}}]{Yang2020}%
  \BibitemOpen
  \bibfield  {author} {\bibinfo {author} {\bibfnamefont {C.~H.}\ \bibnamefont
  {Yang}}, \bibinfo {author} {\bibfnamefont {R.~C.~C.}\ \bibnamefont {Leon}},
  \bibinfo {author} {\bibfnamefont {J.~C.~C.}\ \bibnamefont {Hwang}}, \bibinfo
  {author} {\bibfnamefont {A.}~\bibnamefont {Saraiva}}, \bibinfo {author}
  {\bibfnamefont {T.}~\bibnamefont {Tanttu}}, \bibinfo {author} {\bibfnamefont
  {W.}~\bibnamefont {Huang}}, \bibinfo {author} {\bibfnamefont {J.~C.}\
  \bibnamefont {Lemyre}}, \bibinfo {author} {\bibfnamefont {K.~W.}\
  \bibnamefont {Chan}}, \bibinfo {author} {\bibfnamefont {K.~Y.}\ \bibnamefont
  {Tan}}, \bibinfo {author} {\bibfnamefont {F.~E.}\ \bibnamefont {Hudson}},
  \bibinfo {author} {\bibfnamefont {K.~M.}\ \bibnamefont {Itoh}}, \bibinfo
  {author} {\bibfnamefont {A.}~\bibnamefont {Morello}}, \bibinfo {author}
  {\bibfnamefont {M.}~\bibnamefont {Pioro-Ladrière}}, \bibinfo {author}
  {\bibfnamefont {A.}~\bibnamefont {Laucht}},\ and\ \bibinfo {author}
  {\bibfnamefont {A.~S.}\ \bibnamefont {Dzurak}},\ }\href
  {https://doi.org/10.1038/s41586-020-2171-6} {\bibfield  {journal} {\bibinfo
  {journal} {Nature 2020 580:7803}\ }\textbf {\bibinfo {volume} {580}},\
  \bibinfo {pages} {350} (\bibinfo {year} {2020})},\ \bibinfo {note}
  {publisher: Nature Publishing Group}\BibitemShut {NoStop}%
\bibitem [{\citenamefont {Noiri}\ \emph {et~al.}(2022)\citenamefont {Noiri},
  \citenamefont {Takeda}, \citenamefont {Nakajima}, \citenamefont {Kobayashi},
  \citenamefont {Sammak}, \citenamefont {Scappucci},\ and\ \citenamefont
  {Tarucha}}]{noiri_fast_2022}%
  \BibitemOpen
  \bibfield  {author} {\bibinfo {author} {\bibfnamefont {A.}~\bibnamefont
  {Noiri}}, \bibinfo {author} {\bibfnamefont {K.}~\bibnamefont {Takeda}},
  \bibinfo {author} {\bibfnamefont {T.}~\bibnamefont {Nakajima}}, \bibinfo
  {author} {\bibfnamefont {T.}~\bibnamefont {Kobayashi}}, \bibinfo {author}
  {\bibfnamefont {A.}~\bibnamefont {Sammak}}, \bibinfo {author} {\bibfnamefont
  {G.}~\bibnamefont {Scappucci}},\ and\ \bibinfo {author} {\bibfnamefont
  {S.}~\bibnamefont {Tarucha}},\ }\href
  {https://doi.org/10.1038/s41586-021-04182-y} {\bibfield  {journal} {\bibinfo
  {journal} {Nature}\ }\textbf {\bibinfo {volume} {601}},\ \bibinfo {pages}
  {338} (\bibinfo {year} {2022})},\ \bibinfo {note} {number: 7893 Publisher:
  Nature Publishing Group}\BibitemShut {NoStop}%
\bibitem [{\citenamefont {Mills}\ \emph {et~al.}(2022)\citenamefont {Mills},
  \citenamefont {Guinn}, \citenamefont {Gullans}, \citenamefont {Sigillito},
  \citenamefont {Feldman}, \citenamefont {Nielsen},\ and\ \citenamefont
  {Petta}}]{mills_two-qubit_2022}%
  \BibitemOpen
  \bibfield  {author} {\bibinfo {author} {\bibfnamefont {A.~R.}\ \bibnamefont
  {Mills}}, \bibinfo {author} {\bibfnamefont {C.~R.}\ \bibnamefont {Guinn}},
  \bibinfo {author} {\bibfnamefont {M.~J.}\ \bibnamefont {Gullans}}, \bibinfo
  {author} {\bibfnamefont {A.~J.}\ \bibnamefont {Sigillito}}, \bibinfo {author}
  {\bibfnamefont {M.~M.}\ \bibnamefont {Feldman}}, \bibinfo {author}
  {\bibfnamefont {E.}~\bibnamefont {Nielsen}},\ and\ \bibinfo {author}
  {\bibfnamefont {J.~R.}\ \bibnamefont {Petta}},\ }\href
  {https://doi.org/10.1126/sciadv.abn5130} {\bibfield  {journal} {\bibinfo
  {journal} {Science Advances}\ }\textbf {\bibinfo {volume} {8}},\ \bibinfo
  {pages} {eabn5130} (\bibinfo {year} {2022})},\ \bibinfo {note} {publisher:
  American Association for the Advancement of Science}\BibitemShut {NoStop}%
\bibitem [{\citenamefont {Camenzind}\ \emph {et~al.}(2022)\citenamefont
  {Camenzind}, \citenamefont {Geyer}, \citenamefont {Fuhrer}, \citenamefont
  {Warburton}, \citenamefont {Zumbühl},\ and\ \citenamefont
  {Kuhlmann}}]{camenzind_hole_2022}%
  \BibitemOpen
  \bibfield  {author} {\bibinfo {author} {\bibfnamefont {L.~C.}\ \bibnamefont
  {Camenzind}}, \bibinfo {author} {\bibfnamefont {S.}~\bibnamefont {Geyer}},
  \bibinfo {author} {\bibfnamefont {A.}~\bibnamefont {Fuhrer}}, \bibinfo
  {author} {\bibfnamefont {R.~J.}\ \bibnamefont {Warburton}}, \bibinfo {author}
  {\bibfnamefont {D.~M.}\ \bibnamefont {Zumbühl}},\ and\ \bibinfo {author}
  {\bibfnamefont {A.~V.}\ \bibnamefont {Kuhlmann}},\ }\href
  {https://doi.org/10.1038/s41928-022-00722-0} {\bibfield  {journal} {\bibinfo
  {journal} {Nature Electronics}\ }\textbf {\bibinfo {volume} {5}},\ \bibinfo
  {pages} {178} (\bibinfo {year} {2022})},\ \bibinfo {note} {number: 3
  Publisher: Nature Publishing Group}\BibitemShut {NoStop}%
\bibitem [{\citenamefont {Xue}\ \emph {et~al.}(2022)\citenamefont {Xue},
  \citenamefont {Russ}, \citenamefont {Samkharadze}, \citenamefont {Undseth},
  \citenamefont {Sammak}, \citenamefont {Scappucci},\ and\ \citenamefont
  {Vandersypen}}]{xue_quantum_2022}%
  \BibitemOpen
  \bibfield  {author} {\bibinfo {author} {\bibfnamefont {X.}~\bibnamefont
  {Xue}}, \bibinfo {author} {\bibfnamefont {M.}~\bibnamefont {Russ}}, \bibinfo
  {author} {\bibfnamefont {N.}~\bibnamefont {Samkharadze}}, \bibinfo {author}
  {\bibfnamefont {B.}~\bibnamefont {Undseth}}, \bibinfo {author} {\bibfnamefont
  {A.}~\bibnamefont {Sammak}}, \bibinfo {author} {\bibfnamefont
  {G.}~\bibnamefont {Scappucci}},\ and\ \bibinfo {author} {\bibfnamefont
  {L.~M.~K.}\ \bibnamefont {Vandersypen}},\ }\href
  {https://doi.org/10.1038/s41586-021-04273-w} {\bibfield  {journal} {\bibinfo
  {journal} {Nature}\ }\textbf {\bibinfo {volume} {601}},\ \bibinfo {pages}
  {343} (\bibinfo {year} {2022})}\BibitemShut {NoStop}%
\bibitem [{\citenamefont {Philips}\ \emph {et~al.}(2022)\citenamefont
  {Philips}, \citenamefont {Mądzik}, \citenamefont {Amitonov}, \citenamefont
  {de~Snoo}, \citenamefont {Russ}, \citenamefont {Kalhor}, \citenamefont
  {Volk}, \citenamefont {Lawrie}, \citenamefont {Brousse}, \citenamefont
  {Tryputen}, \citenamefont {Wuetz}, \citenamefont {Sammak}, \citenamefont
  {Veldhorst}, \citenamefont {Scappucci},\ and\ \citenamefont
  {Vandersypen}}]{philips_universal_2022}%
  \BibitemOpen
  \bibfield  {author} {\bibinfo {author} {\bibfnamefont {S.~G.~J.}\
  \bibnamefont {Philips}}, \bibinfo {author} {\bibfnamefont {M.~T.}\
  \bibnamefont {Mądzik}}, \bibinfo {author} {\bibfnamefont {S.~V.}\
  \bibnamefont {Amitonov}}, \bibinfo {author} {\bibfnamefont {S.~L.}\
  \bibnamefont {de~Snoo}}, \bibinfo {author} {\bibfnamefont {M.}~\bibnamefont
  {Russ}}, \bibinfo {author} {\bibfnamefont {N.}~\bibnamefont {Kalhor}},
  \bibinfo {author} {\bibfnamefont {C.}~\bibnamefont {Volk}}, \bibinfo {author}
  {\bibfnamefont {W.~I.~L.}\ \bibnamefont {Lawrie}}, \bibinfo {author}
  {\bibfnamefont {D.}~\bibnamefont {Brousse}}, \bibinfo {author} {\bibfnamefont
  {L.}~\bibnamefont {Tryputen}}, \bibinfo {author} {\bibfnamefont {B.~P.}\
  \bibnamefont {Wuetz}}, \bibinfo {author} {\bibfnamefont {A.}~\bibnamefont
  {Sammak}}, \bibinfo {author} {\bibfnamefont {M.}~\bibnamefont {Veldhorst}},
  \bibinfo {author} {\bibfnamefont {G.}~\bibnamefont {Scappucci}},\ and\
  \bibinfo {author} {\bibfnamefont {L.~M.~K.}\ \bibnamefont {Vandersypen}},\
  }\href {https://doi.org/10.1038/s41586-022-05117-x} {\bibfield  {journal}
  {\bibinfo  {journal} {Nature}\ }\textbf {\bibinfo {volume} {609}},\ \bibinfo
  {pages} {919} (\bibinfo {year} {2022})},\ \bibinfo {note} {number: 7929
  Publisher: Nature Publishing Group}\BibitemShut {NoStop}%
\bibitem [{\citenamefont {Hendrickx}\ \emph {et~al.}(2021)\citenamefont
  {Hendrickx}, \citenamefont {Lawrie}, \citenamefont {Russ}, \citenamefont {van
  Riggelen}, \citenamefont {de~Snoo}, \citenamefont {Schouten}, \citenamefont
  {Sammak}, \citenamefont {Scappucci},\ and\ \citenamefont
  {Veldhorst}}]{hendrickx_four-qubit_2021}%
  \BibitemOpen
  \bibfield  {author} {\bibinfo {author} {\bibfnamefont {N.~W.}\ \bibnamefont
  {Hendrickx}}, \bibinfo {author} {\bibfnamefont {W.~I.~L.}\ \bibnamefont
  {Lawrie}}, \bibinfo {author} {\bibfnamefont {M.}~\bibnamefont {Russ}},
  \bibinfo {author} {\bibfnamefont {F.}~\bibnamefont {van Riggelen}}, \bibinfo
  {author} {\bibfnamefont {S.~L.}\ \bibnamefont {de~Snoo}}, \bibinfo {author}
  {\bibfnamefont {R.~N.}\ \bibnamefont {Schouten}}, \bibinfo {author}
  {\bibfnamefont {A.}~\bibnamefont {Sammak}}, \bibinfo {author} {\bibfnamefont
  {G.}~\bibnamefont {Scappucci}},\ and\ \bibinfo {author} {\bibfnamefont
  {M.}~\bibnamefont {Veldhorst}},\ }\href
  {https://doi.org/10.1038/s41586-021-03332-6} {\bibfield  {journal} {\bibinfo
  {journal} {Nature}\ }\textbf {\bibinfo {volume} {591}},\ \bibinfo {pages}
  {580} (\bibinfo {year} {2021})},\ \bibinfo {note} {number: 7851 Publisher:
  Nature Publishing Group}\BibitemShut {NoStop}%
\bibitem [{\citenamefont {Zwerver}\ \emph {et~al.}(2022)\citenamefont
  {Zwerver}, \citenamefont {Krähenmann}, \citenamefont {Watson}, \citenamefont
  {Lampert}, \citenamefont {George}, \citenamefont {Pillarisetty},
  \citenamefont {Bojarski}, \citenamefont {Amin}, \citenamefont {Amitonov},
  \citenamefont {Boter}, \citenamefont {Caudillo}, \citenamefont
  {Correas-Serrano}, \citenamefont {Dehollain}, \citenamefont {Droulers},
  \citenamefont {Henry}, \citenamefont {Kotlyar}, \citenamefont {Lodari},
  \citenamefont {Lüthi}, \citenamefont {Michalak}, \citenamefont {Mueller},
  \citenamefont {Neyens}, \citenamefont {Roberts}, \citenamefont {Samkharadze},
  \citenamefont {Zheng}, \citenamefont {Zietz}, \citenamefont {Scappucci},
  \citenamefont {Veldhorst}, \citenamefont {Vandersypen},\ and\ \citenamefont
  {Clarke}}]{zwerver_qubits_2022}%
  \BibitemOpen
  \bibfield  {author} {\bibinfo {author} {\bibfnamefont {A.~M.~J.}\
  \bibnamefont {Zwerver}}, \bibinfo {author} {\bibfnamefont {T.}~\bibnamefont
  {Krähenmann}}, \bibinfo {author} {\bibfnamefont {T.~F.}\ \bibnamefont
  {Watson}}, \bibinfo {author} {\bibfnamefont {L.}~\bibnamefont {Lampert}},
  \bibinfo {author} {\bibfnamefont {H.~C.}\ \bibnamefont {George}}, \bibinfo
  {author} {\bibfnamefont {R.}~\bibnamefont {Pillarisetty}}, \bibinfo {author}
  {\bibfnamefont {S.~A.}\ \bibnamefont {Bojarski}}, \bibinfo {author}
  {\bibfnamefont {P.}~\bibnamefont {Amin}}, \bibinfo {author} {\bibfnamefont
  {S.~V.}\ \bibnamefont {Amitonov}}, \bibinfo {author} {\bibfnamefont {J.~M.}\
  \bibnamefont {Boter}}, \bibinfo {author} {\bibfnamefont {R.}~\bibnamefont
  {Caudillo}}, \bibinfo {author} {\bibfnamefont {D.}~\bibnamefont
  {Correas-Serrano}}, \bibinfo {author} {\bibfnamefont {J.~P.}\ \bibnamefont
  {Dehollain}}, \bibinfo {author} {\bibfnamefont {G.}~\bibnamefont {Droulers}},
  \bibinfo {author} {\bibfnamefont {E.~M.}\ \bibnamefont {Henry}}, \bibinfo
  {author} {\bibfnamefont {R.}~\bibnamefont {Kotlyar}}, \bibinfo {author}
  {\bibfnamefont {M.}~\bibnamefont {Lodari}}, \bibinfo {author} {\bibfnamefont
  {F.}~\bibnamefont {Lüthi}}, \bibinfo {author} {\bibfnamefont {D.~J.}\
  \bibnamefont {Michalak}}, \bibinfo {author} {\bibfnamefont {B.~K.}\
  \bibnamefont {Mueller}}, \bibinfo {author} {\bibfnamefont {S.}~\bibnamefont
  {Neyens}}, \bibinfo {author} {\bibfnamefont {J.}~\bibnamefont {Roberts}},
  \bibinfo {author} {\bibfnamefont {N.}~\bibnamefont {Samkharadze}}, \bibinfo
  {author} {\bibfnamefont {G.}~\bibnamefont {Zheng}}, \bibinfo {author}
  {\bibfnamefont {O.~K.}\ \bibnamefont {Zietz}}, \bibinfo {author}
  {\bibfnamefont {G.}~\bibnamefont {Scappucci}}, \bibinfo {author}
  {\bibfnamefont {M.}~\bibnamefont {Veldhorst}}, \bibinfo {author}
  {\bibfnamefont {L.~M.~K.}\ \bibnamefont {Vandersypen}},\ and\ \bibinfo
  {author} {\bibfnamefont {J.~S.}\ \bibnamefont {Clarke}},\ }\href
  {https://doi.org/10.1038/s41928-022-00727-9} {\bibfield  {journal} {\bibinfo
  {journal} {Nature Electronics}\ }\textbf {\bibinfo {volume} {5}},\ \bibinfo
  {pages} {184} (\bibinfo {year} {2022})},\ \bibinfo {note} {number: 3
  Publisher: Nature Publishing Group}\BibitemShut {NoStop}%
\bibitem [{\citenamefont {Elsayed}\ \emph {et~al.}(2022)\citenamefont
  {Elsayed}, \citenamefont {Shehata}, \citenamefont {Godfrin}, \citenamefont
  {Kubicek}, \citenamefont {Massar}, \citenamefont {Canvel}, \citenamefont
  {Jussot}, \citenamefont {Simion}, \citenamefont {Mongillo}, \citenamefont
  {Wan}, \citenamefont {Govoreanu}, \citenamefont {Radu}, \citenamefont {Li},
  \citenamefont {Van~Dorpe},\ and\ \citenamefont
  {De~Greve}}]{elsayed_low_2022}%
  \BibitemOpen
  \bibfield  {author} {\bibinfo {author} {\bibfnamefont {A.}~\bibnamefont
  {Elsayed}}, \bibinfo {author} {\bibfnamefont {M.}~\bibnamefont {Shehata}},
  \bibinfo {author} {\bibfnamefont {C.}~\bibnamefont {Godfrin}}, \bibinfo
  {author} {\bibfnamefont {S.}~\bibnamefont {Kubicek}}, \bibinfo {author}
  {\bibfnamefont {S.}~\bibnamefont {Massar}}, \bibinfo {author} {\bibfnamefont
  {Y.}~\bibnamefont {Canvel}}, \bibinfo {author} {\bibfnamefont
  {J.}~\bibnamefont {Jussot}}, \bibinfo {author} {\bibfnamefont
  {G.}~\bibnamefont {Simion}}, \bibinfo {author} {\bibfnamefont
  {M.}~\bibnamefont {Mongillo}}, \bibinfo {author} {\bibfnamefont
  {D.}~\bibnamefont {Wan}}, \bibinfo {author} {\bibfnamefont {B.}~\bibnamefont
  {Govoreanu}}, \bibinfo {author} {\bibfnamefont {I.~P.}\ \bibnamefont {Radu}},
  \bibinfo {author} {\bibfnamefont {R.}~\bibnamefont {Li}}, \bibinfo {author}
  {\bibfnamefont {P.}~\bibnamefont {Van~Dorpe}},\ and\ \bibinfo {author}
  {\bibfnamefont {K.}~\bibnamefont {De~Greve}},\ }\href
  {https://doi.org/10.48550/arXiv.2212.06464} {\bibinfo {title} {Low charge
  noise quantum dots with industrial {CMOS} manufacturing}} (\bibinfo {year}
  {2022}),\ \bibinfo {note} {arXiv:2212.06464 [cond-mat,
  physics:quant-ph]}\BibitemShut {NoStop}%
\bibitem [{\citenamefont {Neyens}\ \emph {et~al.}(2023)\citenamefont {Neyens},
  \citenamefont {Zietz}, \citenamefont {Watson}, \citenamefont {Luthi},
  \citenamefont {Nethwewala}, \citenamefont {George}, \citenamefont {Henry},
  \citenamefont {Wagner}, \citenamefont {Islam}, \citenamefont {Pillarisetty},
  \citenamefont {Kotlyar}, \citenamefont {Millard}, \citenamefont {Pellerano},
  \citenamefont {Bishop}, \citenamefont {Bojarski}, \citenamefont {Roberts},\
  and\ \citenamefont {Clarke}}]{neyens2023probing}%
  \BibitemOpen
  \bibfield  {author} {\bibinfo {author} {\bibfnamefont {S.}~\bibnamefont
  {Neyens}}, \bibinfo {author} {\bibfnamefont {O.}~\bibnamefont {Zietz}},
  \bibinfo {author} {\bibfnamefont {T.}~\bibnamefont {Watson}}, \bibinfo
  {author} {\bibfnamefont {F.}~\bibnamefont {Luthi}}, \bibinfo {author}
  {\bibfnamefont {A.}~\bibnamefont {Nethwewala}}, \bibinfo {author}
  {\bibfnamefont {H.}~\bibnamefont {George}}, \bibinfo {author} {\bibfnamefont
  {E.}~\bibnamefont {Henry}}, \bibinfo {author} {\bibfnamefont
  {A.}~\bibnamefont {Wagner}}, \bibinfo {author} {\bibfnamefont
  {M.}~\bibnamefont {Islam}}, \bibinfo {author} {\bibfnamefont
  {R.}~\bibnamefont {Pillarisetty}}, \bibinfo {author} {\bibfnamefont
  {R.}~\bibnamefont {Kotlyar}}, \bibinfo {author} {\bibfnamefont
  {K.}~\bibnamefont {Millard}}, \bibinfo {author} {\bibfnamefont
  {S.}~\bibnamefont {Pellerano}}, \bibinfo {author} {\bibfnamefont
  {N.}~\bibnamefont {Bishop}}, \bibinfo {author} {\bibfnamefont
  {S.}~\bibnamefont {Bojarski}}, \bibinfo {author} {\bibfnamefont
  {J.}~\bibnamefont {Roberts}},\ and\ \bibinfo {author} {\bibfnamefont {J.~S.}\
  \bibnamefont {Clarke}},\ }\href@noop {} {\bibinfo {title} {Probing single
  electrons across 300 mm spin qubit wafers}} (\bibinfo {year} {2023}),\
  \Eprint {https://arxiv.org/abs/2307.04812} {arXiv:2307.04812 [quant-ph]}
  \BibitemShut {NoStop}%
\bibitem [{\citenamefont {Sabbagh}\ \emph {et~al.}(2019)\citenamefont
  {Sabbagh}, \citenamefont {Thomas}, \citenamefont {Torres}, \citenamefont
  {Pillarisetty}, \citenamefont {Amin}, \citenamefont {George}, \citenamefont
  {Singh}, \citenamefont {Budrevich}, \citenamefont {Robinson}, \citenamefont
  {Merrill}, \citenamefont {Ross}, \citenamefont {Roberts}, \citenamefont
  {Lampert}, \citenamefont {Massa}, \citenamefont {Amitonov}, \citenamefont
  {Boter}, \citenamefont {Droulers}, \citenamefont {Eenink}, \citenamefont {van
  Hezel}, \citenamefont {Donelson}, \citenamefont {Veldhorst}, \citenamefont
  {Vandersypen}, \citenamefont {Clarke},\ and\ \citenamefont
  {Scappucci}}]{Sabbagh2019}%
  \BibitemOpen
  \bibfield  {author} {\bibinfo {author} {\bibfnamefont {D.}~\bibnamefont
  {Sabbagh}}, \bibinfo {author} {\bibfnamefont {N.}~\bibnamefont {Thomas}},
  \bibinfo {author} {\bibfnamefont {J.}~\bibnamefont {Torres}}, \bibinfo
  {author} {\bibfnamefont {R.}~\bibnamefont {Pillarisetty}}, \bibinfo {author}
  {\bibfnamefont {P.}~\bibnamefont {Amin}}, \bibinfo {author} {\bibfnamefont
  {H.}~\bibnamefont {George}}, \bibinfo {author} {\bibfnamefont
  {K.}~\bibnamefont {Singh}}, \bibinfo {author} {\bibfnamefont
  {A.}~\bibnamefont {Budrevich}}, \bibinfo {author} {\bibfnamefont
  {M.}~\bibnamefont {Robinson}}, \bibinfo {author} {\bibfnamefont
  {D.}~\bibnamefont {Merrill}}, \bibinfo {author} {\bibfnamefont
  {L.}~\bibnamefont {Ross}}, \bibinfo {author} {\bibfnamefont {J.}~\bibnamefont
  {Roberts}}, \bibinfo {author} {\bibfnamefont {L.}~\bibnamefont {Lampert}},
  \bibinfo {author} {\bibfnamefont {L.}~\bibnamefont {Massa}}, \bibinfo
  {author} {\bibfnamefont {S.}~\bibnamefont {Amitonov}}, \bibinfo {author}
  {\bibfnamefont {J.}~\bibnamefont {Boter}}, \bibinfo {author} {\bibfnamefont
  {G.}~\bibnamefont {Droulers}}, \bibinfo {author} {\bibfnamefont
  {H.}~\bibnamefont {Eenink}}, \bibinfo {author} {\bibfnamefont
  {M.}~\bibnamefont {van Hezel}}, \bibinfo {author} {\bibfnamefont
  {D.}~\bibnamefont {Donelson}}, \bibinfo {author} {\bibfnamefont
  {M.}~\bibnamefont {Veldhorst}}, \bibinfo {author} {\bibfnamefont
  {L.}~\bibnamefont {Vandersypen}}, \bibinfo {author} {\bibfnamefont
  {J.}~\bibnamefont {Clarke}},\ and\ \bibinfo {author} {\bibfnamefont
  {G.}~\bibnamefont {Scappucci}},\ }\href
  {https://doi.org/10.1103/PhysRevApplied.12.014013} {\bibfield  {journal}
  {\bibinfo  {journal} {Phys. Rev. Appl.}\ }\textbf {\bibinfo {volume} {12}},\
  \bibinfo {pages} {014013} (\bibinfo {year} {2019})}\BibitemShut {NoStop}%
\bibitem [{\citenamefont {Gidney}\ and\ \citenamefont
  {Ekerå}(2021)}]{gidney_how_2021}%
  \BibitemOpen
  \bibfield  {author} {\bibinfo {author} {\bibfnamefont {C.}~\bibnamefont
  {Gidney}}\ and\ \bibinfo {author} {\bibfnamefont {M.}~\bibnamefont
  {Ekerå}},\ }\href {https://doi.org/10.22331/q-2021-04-15-433} {\bibfield
  {journal} {\bibinfo  {journal} {Quantum}\ }\textbf {\bibinfo {volume} {5}},\
  \bibinfo {pages} {433} (\bibinfo {year} {2021})},\ \bibinfo {note}
  {publisher: Verein zur Förderung des Open Access Publizierens in den
  Quantenwissenschaften}\BibitemShut {NoStop}%
\bibitem [{\citenamefont {Beverland}\ \emph {et~al.}(2022)\citenamefont
  {Beverland}, \citenamefont {Murali}, \citenamefont {Troyer}, \citenamefont
  {Svore}, \citenamefont {Hoefler}, \citenamefont {Kliuchnikov}, \citenamefont
  {Low}, \citenamefont {Soeken}, \citenamefont {Sundaram},\ and\ \citenamefont
  {Vaschillo}}]{beverland_assessing_2022}%
  \BibitemOpen
  \bibfield  {author} {\bibinfo {author} {\bibfnamefont {M.~E.}\ \bibnamefont
  {Beverland}}, \bibinfo {author} {\bibfnamefont {P.}~\bibnamefont {Murali}},
  \bibinfo {author} {\bibfnamefont {M.}~\bibnamefont {Troyer}}, \bibinfo
  {author} {\bibfnamefont {K.~M.}\ \bibnamefont {Svore}}, \bibinfo {author}
  {\bibfnamefont {T.}~\bibnamefont {Hoefler}}, \bibinfo {author} {\bibfnamefont
  {V.}~\bibnamefont {Kliuchnikov}}, \bibinfo {author} {\bibfnamefont {G.~H.}\
  \bibnamefont {Low}}, \bibinfo {author} {\bibfnamefont {M.}~\bibnamefont
  {Soeken}}, \bibinfo {author} {\bibfnamefont {A.}~\bibnamefont {Sundaram}},\
  and\ \bibinfo {author} {\bibfnamefont {A.}~\bibnamefont {Vaschillo}},\ }\href
  {https://doi.org/10.48550/arXiv.2211.07629} {\bibinfo {title} {Assessing
  requirements to scale to practical quantum advantage}} (\bibinfo {year}
  {2022}),\ \bibinfo {note} {arXiv:2211.07629 [quant-ph]}\BibitemShut {NoStop}%
\bibitem [{\citenamefont {Tanttu}\ \emph {et~al.}(2019)\citenamefont {Tanttu},
  \citenamefont {Hensen}, \citenamefont {Chan}, \citenamefont {Yang},
  \citenamefont {Huang}, \citenamefont {Fogarty}, \citenamefont {Hudson},
  \citenamefont {Itoh}, \citenamefont {Culcer}, \citenamefont {Laucht},
  \citenamefont {Morello},\ and\ \citenamefont {Dzurak}}]{Tanttu2019}%
  \BibitemOpen
  \bibfield  {author} {\bibinfo {author} {\bibfnamefont {T.}~\bibnamefont
  {Tanttu}}, \bibinfo {author} {\bibfnamefont {B.}~\bibnamefont {Hensen}},
  \bibinfo {author} {\bibfnamefont {K.~W.}\ \bibnamefont {Chan}}, \bibinfo
  {author} {\bibfnamefont {C.~H.}\ \bibnamefont {Yang}}, \bibinfo {author}
  {\bibfnamefont {W.~W.}\ \bibnamefont {Huang}}, \bibinfo {author}
  {\bibfnamefont {M.}~\bibnamefont {Fogarty}}, \bibinfo {author} {\bibfnamefont
  {F.}~\bibnamefont {Hudson}}, \bibinfo {author} {\bibfnamefont
  {K.}~\bibnamefont {Itoh}}, \bibinfo {author} {\bibfnamefont {D.}~\bibnamefont
  {Culcer}}, \bibinfo {author} {\bibfnamefont {A.}~\bibnamefont {Laucht}},
  \bibinfo {author} {\bibfnamefont {A.}~\bibnamefont {Morello}},\ and\ \bibinfo
  {author} {\bibfnamefont {A.}~\bibnamefont {Dzurak}},\ }\bibfield  {journal}
  {\bibinfo  {journal} {Physical Review X}\ }\textbf {\bibinfo {volume} {9}},\
  \href {https://doi.org/10.1103/PhysRevX.9.021028} {10.1103/PhysRevX.9.021028}
  (\bibinfo {year} {2019}),\ \bibinfo {note} {arXiv: 1807.10415 Publisher:
  American Physical Society (APS)}\BibitemShut {NoStop}%
\bibitem [{\citenamefont {Saraiva}\ \emph {et~al.}(2022)\citenamefont
  {Saraiva}, \citenamefont {Lim}, \citenamefont {Yang}, \citenamefont {Escott},
  \citenamefont {Laucht},\ and\ \citenamefont {Dzurak}}]{Saraiva2022}%
  \BibitemOpen
  \bibfield  {author} {\bibinfo {author} {\bibfnamefont {A.}~\bibnamefont
  {Saraiva}}, \bibinfo {author} {\bibfnamefont {W.~H.}\ \bibnamefont {Lim}},
  \bibinfo {author} {\bibfnamefont {C.~H.}\ \bibnamefont {Yang}}, \bibinfo
  {author} {\bibfnamefont {C.~C.}\ \bibnamefont {Escott}}, \bibinfo {author}
  {\bibfnamefont {A.}~\bibnamefont {Laucht}},\ and\ \bibinfo {author}
  {\bibfnamefont {A.~S.}\ \bibnamefont {Dzurak}},\ }\href
  {https://doi.org/10.1002/adfm.202105488} {\bibfield  {journal} {\bibinfo
  {journal} {Advanced Functional Materials}\ }\textbf {\bibinfo {volume}
  {32}},\ \bibinfo {pages} {2105488} (\bibinfo {year} {2022})},\ \bibinfo
  {note} {publisher: John Wiley and Sons Inc}\BibitemShut {NoStop}%
\bibitem [{\citenamefont {Cifuentes}\ \emph {et~al.}(2023)\citenamefont
  {Cifuentes}, \citenamefont {Tanttu}, \citenamefont {Gilbert}, \citenamefont
  {Huang}, \citenamefont {Vahapoglu}, \citenamefont {Leon}, \citenamefont
  {Serrano}, \citenamefont {Otter}, \citenamefont {Dunmore}, \citenamefont
  {Mai}, \citenamefont {Schlattner}, \citenamefont {Feng}, \citenamefont
  {Itoh}, \citenamefont {Abrosimov}, \citenamefont {Pohl}, \citenamefont
  {Thewalt}, \citenamefont {Laucht}, \citenamefont {Yang}, \citenamefont
  {Escott}, \citenamefont {Lim}, \citenamefont {Hudson}, \citenamefont
  {Rahman}, \citenamefont {Saraiva},\ and\ \citenamefont
  {Dzurak}}]{cifuentes_bounds_2023}%
  \BibitemOpen
  \bibfield  {author} {\bibinfo {author} {\bibfnamefont {J.~D.}\ \bibnamefont
  {Cifuentes}}, \bibinfo {author} {\bibfnamefont {T.}~\bibnamefont {Tanttu}},
  \bibinfo {author} {\bibfnamefont {W.}~\bibnamefont {Gilbert}}, \bibinfo
  {author} {\bibfnamefont {J.~Y.}\ \bibnamefont {Huang}}, \bibinfo {author}
  {\bibfnamefont {E.}~\bibnamefont {Vahapoglu}}, \bibinfo {author}
  {\bibfnamefont {R.~C.~C.}\ \bibnamefont {Leon}}, \bibinfo {author}
  {\bibfnamefont {S.}~\bibnamefont {Serrano}}, \bibinfo {author} {\bibfnamefont
  {D.}~\bibnamefont {Otter}}, \bibinfo {author} {\bibfnamefont
  {D.}~\bibnamefont {Dunmore}}, \bibinfo {author} {\bibfnamefont {P.~Y.}\
  \bibnamefont {Mai}}, \bibinfo {author} {\bibfnamefont {F.}~\bibnamefont
  {Schlattner}}, \bibinfo {author} {\bibfnamefont {M.}~\bibnamefont {Feng}},
  \bibinfo {author} {\bibfnamefont {K.}~\bibnamefont {Itoh}}, \bibinfo {author}
  {\bibfnamefont {N.}~\bibnamefont {Abrosimov}}, \bibinfo {author}
  {\bibfnamefont {H.-J.}\ \bibnamefont {Pohl}}, \bibinfo {author}
  {\bibfnamefont {M.}~\bibnamefont {Thewalt}}, \bibinfo {author} {\bibfnamefont
  {A.}~\bibnamefont {Laucht}}, \bibinfo {author} {\bibfnamefont {C.~H.}\
  \bibnamefont {Yang}}, \bibinfo {author} {\bibfnamefont {C.~C.}\ \bibnamefont
  {Escott}}, \bibinfo {author} {\bibfnamefont {W.~H.}\ \bibnamefont {Lim}},
  \bibinfo {author} {\bibfnamefont {F.~E.}\ \bibnamefont {Hudson}}, \bibinfo
  {author} {\bibfnamefont {R.}~\bibnamefont {Rahman}}, \bibinfo {author}
  {\bibfnamefont {A.}~\bibnamefont {Saraiva}},\ and\ \bibinfo {author}
  {\bibfnamefont {A.~S.}\ \bibnamefont {Dzurak}},\ }\href
  {https://doi.org/10.48550/arXiv.2303.14864} {\bibinfo {title} {Bounds to
  electron spin qubit variability for scalable {CMOS} architectures}} (\bibinfo
  {year} {2023}),\ \bibinfo {note} {arXiv:2303.14864 [cond-mat,
  physics:quant-ph]}\BibitemShut {NoStop}%
\bibitem [{\citenamefont {Hansen}\ \emph {et~al.}(2021)\citenamefont {Hansen},
  \citenamefont {Seedhouse}, \citenamefont {Saraiva}, \citenamefont {Laucht},
  \citenamefont {Dzurak},\ and\ \citenamefont {Yang}}]{hansen_pulse_2021}%
  \BibitemOpen
  \bibfield  {author} {\bibinfo {author} {\bibfnamefont {I.}~\bibnamefont
  {Hansen}}, \bibinfo {author} {\bibfnamefont {A.~E.}\ \bibnamefont
  {Seedhouse}}, \bibinfo {author} {\bibfnamefont {A.}~\bibnamefont {Saraiva}},
  \bibinfo {author} {\bibfnamefont {A.}~\bibnamefont {Laucht}}, \bibinfo
  {author} {\bibfnamefont {A.~S.}\ \bibnamefont {Dzurak}},\ and\ \bibinfo
  {author} {\bibfnamefont {C.~H.}\ \bibnamefont {Yang}},\ }\href
  {https://doi.org/10.1103/PhysRevA.104.062415} {\bibfield  {journal} {\bibinfo
   {journal} {Physical Review A}\ }\textbf {\bibinfo {volume} {104}},\ \bibinfo
  {pages} {062415} (\bibinfo {year} {2021})},\ \bibinfo {note} {publisher:
  American Physical Society}\BibitemShut {NoStop}%
\bibitem [{\citenamefont {Vahapoglu}\ \emph {et~al.}(2020)\citenamefont
  {Vahapoglu}, \citenamefont {Slack-Smith}, \citenamefont {Leon}, \citenamefont
  {Lim}, \citenamefont {Hudson}, \citenamefont {Day}, \citenamefont {Tanttu},
  \citenamefont {Yang}, \citenamefont {Laucht}, \citenamefont {Dzurak},\ and\
  \citenamefont {Pla}}]{Vahapoglu2020}%
  \BibitemOpen
  \bibfield  {author} {\bibinfo {author} {\bibfnamefont {E.}~\bibnamefont
  {Vahapoglu}}, \bibinfo {author} {\bibfnamefont {J.~P.}\ \bibnamefont
  {Slack-Smith}}, \bibinfo {author} {\bibfnamefont {R.~C.~C.}\ \bibnamefont
  {Leon}}, \bibinfo {author} {\bibfnamefont {W.~H.}\ \bibnamefont {Lim}},
  \bibinfo {author} {\bibfnamefont {F.~E.}\ \bibnamefont {Hudson}}, \bibinfo
  {author} {\bibfnamefont {T.}~\bibnamefont {Day}}, \bibinfo {author}
  {\bibfnamefont {T.}~\bibnamefont {Tanttu}}, \bibinfo {author} {\bibfnamefont
  {C.~H.}\ \bibnamefont {Yang}}, \bibinfo {author} {\bibfnamefont
  {A.}~\bibnamefont {Laucht}}, \bibinfo {author} {\bibfnamefont {A.~S.}\
  \bibnamefont {Dzurak}},\ and\ \bibinfo {author} {\bibfnamefont {J.~J.}\
  \bibnamefont {Pla}},\ }\bibfield  {journal} {\bibinfo  {journal} {Science
  Advances}\ }\textbf {\bibinfo {volume} {7}},\ \href
  {https://doi.org/10.1126/sciadv.abg9158} {10.1126/sciadv.abg9158} (\bibinfo
  {year} {2020}),\ \bibinfo {note} {arXiv: 2012.10225 Publisher: American
  Association for the Advancement of Science}\BibitemShut {NoStop}%
\bibitem [{\citenamefont {Seedhouse}\ \emph {et~al.}(2021)\citenamefont
  {Seedhouse}, \citenamefont {Hansen}, \citenamefont {Laucht}, \citenamefont
  {Yang}, \citenamefont {Dzurak},\ and\ \citenamefont
  {Saraiva}}]{seedhouse_quantum_2021}%
  \BibitemOpen
  \bibfield  {author} {\bibinfo {author} {\bibfnamefont {A.~E.}\ \bibnamefont
  {Seedhouse}}, \bibinfo {author} {\bibfnamefont {I.}~\bibnamefont {Hansen}},
  \bibinfo {author} {\bibfnamefont {A.}~\bibnamefont {Laucht}}, \bibinfo
  {author} {\bibfnamefont {C.~H.}\ \bibnamefont {Yang}}, \bibinfo {author}
  {\bibfnamefont {A.~S.}\ \bibnamefont {Dzurak}},\ and\ \bibinfo {author}
  {\bibfnamefont {A.}~\bibnamefont {Saraiva}},\ }\href
  {https://doi.org/10.1103/PhysRevB.104.235411} {\bibfield  {journal} {\bibinfo
   {journal} {Physical Review B}\ }\textbf {\bibinfo {volume} {104}},\ \bibinfo
  {pages} {235411} (\bibinfo {year} {2021})},\ \bibinfo {note} {publisher:
  American Physical Society}\BibitemShut {NoStop}%
\bibitem [{\citenamefont {Hansen}\ \emph {et~al.}(2022)\citenamefont {Hansen},
  \citenamefont {Seedhouse}, \citenamefont {Chan}, \citenamefont {Hudson},
  \citenamefont {Itoh}, \citenamefont {Laucht}, \citenamefont {Saraiva},
  \citenamefont {Yang},\ and\ \citenamefont
  {Dzurak}}]{hansen_implementation_2022}%
  \BibitemOpen
  \bibfield  {author} {\bibinfo {author} {\bibfnamefont {I.}~\bibnamefont
  {Hansen}}, \bibinfo {author} {\bibfnamefont {A.~E.}\ \bibnamefont
  {Seedhouse}}, \bibinfo {author} {\bibfnamefont {K.~W.}\ \bibnamefont {Chan}},
  \bibinfo {author} {\bibfnamefont {F.~E.}\ \bibnamefont {Hudson}}, \bibinfo
  {author} {\bibfnamefont {K.~M.}\ \bibnamefont {Itoh}}, \bibinfo {author}
  {\bibfnamefont {A.}~\bibnamefont {Laucht}}, \bibinfo {author} {\bibfnamefont
  {A.}~\bibnamefont {Saraiva}}, \bibinfo {author} {\bibfnamefont {C.~H.}\
  \bibnamefont {Yang}},\ and\ \bibinfo {author} {\bibfnamefont {A.~S.}\
  \bibnamefont {Dzurak}},\ }\href {https://doi.org/10.1063/5.0096467}
  {\bibfield  {journal} {\bibinfo  {journal} {Applied Physics Reviews}\
  }\textbf {\bibinfo {volume} {9}},\ \bibinfo {pages} {031409} (\bibinfo {year}
  {2022})},\ \bibinfo {note} {publisher: American Institute of
  Physics}\BibitemShut {NoStop}%
\bibitem [{\citenamefont {Kane}(1998)}]{kane_silicon-based_1998}%
  \BibitemOpen
  \bibfield  {author} {\bibinfo {author} {\bibfnamefont {B.~E.}\ \bibnamefont
  {Kane}},\ }\href {https://doi.org/10.1038/30156} {\bibfield  {journal}
  {\bibinfo  {journal} {Nature}\ }\textbf {\bibinfo {volume} {393}},\ \bibinfo
  {pages} {133} (\bibinfo {year} {1998})}\BibitemShut {NoStop}%
\bibitem [{\citenamefont {Ferdous}\ \emph {et~al.}(2018)\citenamefont
  {Ferdous}, \citenamefont {Chan}, \citenamefont {Veldhorst}, \citenamefont
  {Hwang}, \citenamefont {Yang}, \citenamefont {Sahasrabudhe}, \citenamefont
  {Klimeck}, \citenamefont {Morello}, \citenamefont {Dzurak},\ and\
  \citenamefont {Rahman}}]{Ferdous2018}%
  \BibitemOpen
  \bibfield  {author} {\bibinfo {author} {\bibfnamefont {R.}~\bibnamefont
  {Ferdous}}, \bibinfo {author} {\bibfnamefont {K.~W.}\ \bibnamefont {Chan}},
  \bibinfo {author} {\bibfnamefont {M.}~\bibnamefont {Veldhorst}}, \bibinfo
  {author} {\bibfnamefont {J.~C.}\ \bibnamefont {Hwang}}, \bibinfo {author}
  {\bibfnamefont {C.~H.}\ \bibnamefont {Yang}}, \bibinfo {author}
  {\bibfnamefont {H.}~\bibnamefont {Sahasrabudhe}}, \bibinfo {author}
  {\bibfnamefont {G.}~\bibnamefont {Klimeck}}, \bibinfo {author} {\bibfnamefont
  {A.}~\bibnamefont {Morello}}, \bibinfo {author} {\bibfnamefont {A.~S.}\
  \bibnamefont {Dzurak}},\ and\ \bibinfo {author} {\bibfnamefont
  {R.}~\bibnamefont {Rahman}},\ }\href
  {https://doi.org/10.1103/PhysRevB.97.241401} {\bibfield  {journal} {\bibinfo
  {journal} {Physical Review B}\ }\textbf {\bibinfo {volume} {97}},\ \bibinfo
  {pages} {241401} (\bibinfo {year} {2018})},\ \bibinfo {note} {arXiv:
  1703.03840 Publisher: American Physical Society}\BibitemShut {NoStop}%
\bibitem [{\citenamefont {Leon}\ \emph {et~al.}(2020)\citenamefont {Leon},
  \citenamefont {Yang}, \citenamefont {Hwang}, \citenamefont {Lemyre},
  \citenamefont {Tanttu}, \citenamefont {Huang}, \citenamefont {Chan},
  \citenamefont {Tan}, \citenamefont {Hudson}, \citenamefont {Itoh},
  \citenamefont {Morello}, \citenamefont {Laucht}, \citenamefont
  {Pioro-Ladrière}, \citenamefont {Saraiva},\ and\ \citenamefont
  {Dzurak}}]{Leon2020}%
  \BibitemOpen
  \bibfield  {author} {\bibinfo {author} {\bibfnamefont {R.~C.}\ \bibnamefont
  {Leon}}, \bibinfo {author} {\bibfnamefont {C.~H.}\ \bibnamefont {Yang}},
  \bibinfo {author} {\bibfnamefont {J.~C.}\ \bibnamefont {Hwang}}, \bibinfo
  {author} {\bibfnamefont {J.~C.}\ \bibnamefont {Lemyre}}, \bibinfo {author}
  {\bibfnamefont {T.}~\bibnamefont {Tanttu}}, \bibinfo {author} {\bibfnamefont
  {W.}~\bibnamefont {Huang}}, \bibinfo {author} {\bibfnamefont {K.~W.}\
  \bibnamefont {Chan}}, \bibinfo {author} {\bibfnamefont {K.~Y.}\ \bibnamefont
  {Tan}}, \bibinfo {author} {\bibfnamefont {F.~E.}\ \bibnamefont {Hudson}},
  \bibinfo {author} {\bibfnamefont {K.~M.}\ \bibnamefont {Itoh}}, \bibinfo
  {author} {\bibfnamefont {A.}~\bibnamefont {Morello}}, \bibinfo {author}
  {\bibfnamefont {A.}~\bibnamefont {Laucht}}, \bibinfo {author} {\bibfnamefont
  {M.}~\bibnamefont {Pioro-Ladrière}}, \bibinfo {author} {\bibfnamefont
  {A.}~\bibnamefont {Saraiva}},\ and\ \bibinfo {author} {\bibfnamefont {A.~S.}\
  \bibnamefont {Dzurak}},\ }\href {https://doi.org/10.1038/s41467-019-14053-w}
  {\bibfield  {journal} {\bibinfo  {journal} {Nature Communications}\ }\textbf
  {\bibinfo {volume} {11}},\ \bibinfo {pages} {1} (\bibinfo {year} {2020})},\
  \bibinfo {note} {arXiv: 1902.01550 Publisher: Nature Research}\BibitemShut
  {NoStop}%
\bibitem [{\citenamefont {Veldhorst}\ \emph {et~al.}(2015)\citenamefont
  {Veldhorst}, \citenamefont {Ruskov}, \citenamefont {Yang}, \citenamefont
  {Hwang}, \citenamefont {Hudson}, \citenamefont {Flatt\'e}, \citenamefont
  {Tahan}, \citenamefont {Itoh}, \citenamefont {Morello},\ and\ \citenamefont
  {Dzurak}}]{Veldhorst2015}%
  \BibitemOpen
  \bibfield  {author} {\bibinfo {author} {\bibfnamefont {M.}~\bibnamefont
  {Veldhorst}}, \bibinfo {author} {\bibfnamefont {R.}~\bibnamefont {Ruskov}},
  \bibinfo {author} {\bibfnamefont {C.~H.}\ \bibnamefont {Yang}}, \bibinfo
  {author} {\bibfnamefont {J.~C.~C.}\ \bibnamefont {Hwang}}, \bibinfo {author}
  {\bibfnamefont {F.~E.}\ \bibnamefont {Hudson}}, \bibinfo {author}
  {\bibfnamefont {M.~E.}\ \bibnamefont {Flatt\'e}}, \bibinfo {author}
  {\bibfnamefont {C.}~\bibnamefont {Tahan}}, \bibinfo {author} {\bibfnamefont
  {K.~M.}\ \bibnamefont {Itoh}}, \bibinfo {author} {\bibfnamefont
  {A.}~\bibnamefont {Morello}},\ and\ \bibinfo {author} {\bibfnamefont {A.~S.}\
  \bibnamefont {Dzurak}},\ }\href {https://doi.org/10.1103/PhysRevB.92.201401}
  {\bibfield  {journal} {\bibinfo  {journal} {Phys. Rev. B}\ }\textbf {\bibinfo
  {volume} {92}},\ \bibinfo {pages} {201401} (\bibinfo {year}
  {2015})}\BibitemShut {NoStop}%
\bibitem [{\citenamefont {Lawrie}\ \emph {et~al.}(2020)\citenamefont {Lawrie},
  \citenamefont {Hendrickx}, \citenamefont {van Riggelen}, \citenamefont
  {Russ}, \citenamefont {Petit}, \citenamefont {Sammak}, \citenamefont
  {Scappucci},\ and\ \citenamefont {Veldhorst}}]{lawrie_spin_2020}%
  \BibitemOpen
  \bibfield  {author} {\bibinfo {author} {\bibfnamefont {W.~I.~L.}\
  \bibnamefont {Lawrie}}, \bibinfo {author} {\bibfnamefont {N.~W.}\
  \bibnamefont {Hendrickx}}, \bibinfo {author} {\bibfnamefont {F.}~\bibnamefont
  {van Riggelen}}, \bibinfo {author} {\bibfnamefont {M.}~\bibnamefont {Russ}},
  \bibinfo {author} {\bibfnamefont {L.}~\bibnamefont {Petit}}, \bibinfo
  {author} {\bibfnamefont {A.}~\bibnamefont {Sammak}}, \bibinfo {author}
  {\bibfnamefont {G.}~\bibnamefont {Scappucci}},\ and\ \bibinfo {author}
  {\bibfnamefont {M.}~\bibnamefont {Veldhorst}},\ }\href
  {https://doi.org/10.1021/acs.nanolett.0c02589} {\bibfield  {journal}
  {\bibinfo  {journal} {Nano Letters}\ }\textbf {\bibinfo {volume} {20}},\
  \bibinfo {pages} {7237} (\bibinfo {year} {2020})},\ \bibinfo {note}
  {publisher: American Chemical Society}\BibitemShut {NoStop}%
\bibitem [{\citenamefont {Hollmann}\ \emph {et~al.}(2020)\citenamefont
  {Hollmann}, \citenamefont {Struck}, \citenamefont {Langrock}, \citenamefont
  {Schmidbauer}, \citenamefont {Schauer}, \citenamefont {Leonhardt},
  \citenamefont {Sawano}, \citenamefont {Riemann}, \citenamefont {Abrosimov},
  \citenamefont {Bougeard},\ and\ \citenamefont
  {Schreiber}}]{hollmann_large_2020}%
  \BibitemOpen
  \bibfield  {author} {\bibinfo {author} {\bibfnamefont {A.}~\bibnamefont
  {Hollmann}}, \bibinfo {author} {\bibfnamefont {T.}~\bibnamefont {Struck}},
  \bibinfo {author} {\bibfnamefont {V.}~\bibnamefont {Langrock}}, \bibinfo
  {author} {\bibfnamefont {A.}~\bibnamefont {Schmidbauer}}, \bibinfo {author}
  {\bibfnamefont {F.}~\bibnamefont {Schauer}}, \bibinfo {author} {\bibfnamefont
  {T.}~\bibnamefont {Leonhardt}}, \bibinfo {author} {\bibfnamefont
  {K.}~\bibnamefont {Sawano}}, \bibinfo {author} {\bibfnamefont
  {H.}~\bibnamefont {Riemann}}, \bibinfo {author} {\bibfnamefont {N.~V.}\
  \bibnamefont {Abrosimov}}, \bibinfo {author} {\bibfnamefont {D.}~\bibnamefont
  {Bougeard}},\ and\ \bibinfo {author} {\bibfnamefont {L.~R.}\ \bibnamefont
  {Schreiber}},\ }\href {https://doi.org/10.1103/PhysRevApplied.13.034068}
  {\bibfield  {journal} {\bibinfo  {journal} {Physical Review Applied}\
  }\textbf {\bibinfo {volume} {13}},\ \bibinfo {pages} {034068} (\bibinfo
  {year} {2020})},\ \bibinfo {note} {publisher: American Physical
  Society}\BibitemShut {NoStop}%
\bibitem [{\citenamefont {Klimeck}\ \emph {et~al.}(2007)\citenamefont
  {Klimeck}, \citenamefont {Ahmed}, \citenamefont {Bae}, \citenamefont {Clark},
  \citenamefont {Haley}, \citenamefont {Lee}, \citenamefont {Naumov},
  \citenamefont {Ryu}, \citenamefont {Saied}, \citenamefont {Prada},
  \citenamefont {Korkusinski}, \citenamefont {Boykin},\ and\ \citenamefont
  {Rahman}}]{Klimeck2007}%
  \BibitemOpen
  \bibfield  {author} {\bibinfo {author} {\bibfnamefont {G.}~\bibnamefont
  {Klimeck}}, \bibinfo {author} {\bibfnamefont {S.~S.}\ \bibnamefont {Ahmed}},
  \bibinfo {author} {\bibfnamefont {H.}~\bibnamefont {Bae}}, \bibinfo {author}
  {\bibfnamefont {S.}~\bibnamefont {Clark}}, \bibinfo {author} {\bibfnamefont
  {B.}~\bibnamefont {Haley}}, \bibinfo {author} {\bibfnamefont
  {S.}~\bibnamefont {Lee}}, \bibinfo {author} {\bibfnamefont {M.}~\bibnamefont
  {Naumov}}, \bibinfo {author} {\bibfnamefont {H.}~\bibnamefont {Ryu}},
  \bibinfo {author} {\bibfnamefont {F.}~\bibnamefont {Saied}}, \bibinfo
  {author} {\bibfnamefont {M.}~\bibnamefont {Prada}}, \bibinfo {author}
  {\bibfnamefont {M.}~\bibnamefont {Korkusinski}}, \bibinfo {author}
  {\bibfnamefont {T.~B.}\ \bibnamefont {Boykin}},\ and\ \bibinfo {author}
  {\bibfnamefont {R.}~\bibnamefont {Rahman}},\ }\href
  {https://doi.org/10.1109/TED.2007.902879} {\bibfield  {journal} {\bibinfo
  {journal} {IEEE Transactions on Electron Devices}\ }\textbf {\bibinfo
  {volume} {54}},\ \bibinfo {pages} {2079} (\bibinfo {year}
  {2007})}\BibitemShut {NoStop}%
\bibitem [{\citenamefont {Naumov}\ \emph {et~al.}(2008)\citenamefont {Naumov},
  \citenamefont {Lee}, \citenamefont {Haley}, \citenamefont {Bae},
  \citenamefont {Clark}, \citenamefont {Rahman}, \citenamefont {Ryu},
  \citenamefont {Saied},\ and\ \citenamefont {Klimeck}}]{Naumov2008}%
  \BibitemOpen
  \bibfield  {author} {\bibinfo {author} {\bibfnamefont {M.}~\bibnamefont
  {Naumov}}, \bibinfo {author} {\bibfnamefont {S.}~\bibnamefont {Lee}},
  \bibinfo {author} {\bibfnamefont {B.}~\bibnamefont {Haley}}, \bibinfo
  {author} {\bibfnamefont {H.}~\bibnamefont {Bae}}, \bibinfo {author}
  {\bibfnamefont {S.}~\bibnamefont {Clark}}, \bibinfo {author} {\bibfnamefont
  {R.}~\bibnamefont {Rahman}}, \bibinfo {author} {\bibfnamefont
  {H.}~\bibnamefont {Ryu}}, \bibinfo {author} {\bibfnamefont {F.}~\bibnamefont
  {Saied}},\ and\ \bibinfo {author} {\bibfnamefont {G.}~\bibnamefont
  {Klimeck}},\ }\href {https://doi.org/10.1007/s10825-008-0223-5} {\bibfield
  {journal} {\bibinfo  {journal} {Journal of Computational Electronics}\
  }\textbf {\bibinfo {volume} {7}},\ \bibinfo {pages} {297} (\bibinfo {year}
  {2008})},\ \bibinfo {note} {publisher: Springer}\BibitemShut {NoStop}%
\bibitem [{\citenamefont {Ruskov}\ \emph {et~al.}(2018)\citenamefont {Ruskov},
  \citenamefont {Veldhorst}, \citenamefont {Dzurak},\ and\ \citenamefont
  {Tahan}}]{Ruskov2018}%
  \BibitemOpen
  \bibfield  {author} {\bibinfo {author} {\bibfnamefont {R.}~\bibnamefont
  {Ruskov}}, \bibinfo {author} {\bibfnamefont {M.}~\bibnamefont {Veldhorst}},
  \bibinfo {author} {\bibfnamefont {A.~S.}\ \bibnamefont {Dzurak}},\ and\
  \bibinfo {author} {\bibfnamefont {C.}~\bibnamefont {Tahan}},\ }\href
  {https://doi.org/10.1103/PhysRevB.98.245424} {\bibfield  {journal} {\bibinfo
  {journal} {Physical Review B}\ }\textbf {\bibinfo {volume} {98}},\ \bibinfo
  {pages} {245424} (\bibinfo {year} {2018})},\ \bibinfo {note} {arXiv:
  1708.04555 Publisher: American Physical Society}\BibitemShut {NoStop}%
\bibitem [{\citenamefont {Zhao}\ \emph {et~al.}(2019)\citenamefont {Zhao},
  \citenamefont {Tanttu}, \citenamefont {Tan}, \citenamefont {Hensen},
  \citenamefont {Chan}, \citenamefont {Hwang}, \citenamefont {Leon},
  \citenamefont {Yang}, \citenamefont {Gilbert}, \citenamefont {Hudson},
  \citenamefont {Itoh}, \citenamefont {Kiselev}, \citenamefont {Ladd},
  \citenamefont {Morello}, \citenamefont {Laucht},\ and\ \citenamefont
  {Dzurak}}]{zhao_single-spin_2019}%
  \BibitemOpen
  \bibfield  {author} {\bibinfo {author} {\bibfnamefont {R.}~\bibnamefont
  {Zhao}}, \bibinfo {author} {\bibfnamefont {T.}~\bibnamefont {Tanttu}},
  \bibinfo {author} {\bibfnamefont {K.~Y.}\ \bibnamefont {Tan}}, \bibinfo
  {author} {\bibfnamefont {B.}~\bibnamefont {Hensen}}, \bibinfo {author}
  {\bibfnamefont {K.~W.}\ \bibnamefont {Chan}}, \bibinfo {author}
  {\bibfnamefont {J.~C.~C.}\ \bibnamefont {Hwang}}, \bibinfo {author}
  {\bibfnamefont {R.~C.~C.}\ \bibnamefont {Leon}}, \bibinfo {author}
  {\bibfnamefont {C.~H.}\ \bibnamefont {Yang}}, \bibinfo {author}
  {\bibfnamefont {W.}~\bibnamefont {Gilbert}}, \bibinfo {author} {\bibfnamefont
  {F.~E.}\ \bibnamefont {Hudson}}, \bibinfo {author} {\bibfnamefont {K.~M.}\
  \bibnamefont {Itoh}}, \bibinfo {author} {\bibfnamefont {A.~A.}\ \bibnamefont
  {Kiselev}}, \bibinfo {author} {\bibfnamefont {T.~D.}\ \bibnamefont {Ladd}},
  \bibinfo {author} {\bibfnamefont {A.}~\bibnamefont {Morello}}, \bibinfo
  {author} {\bibfnamefont {A.}~\bibnamefont {Laucht}},\ and\ \bibinfo {author}
  {\bibfnamefont {A.~S.}\ \bibnamefont {Dzurak}},\ }\href
  {https://doi.org/10.1038/s41467-019-13416-7} {\bibfield  {journal} {\bibinfo
  {journal} {Nature Communications}\ }\textbf {\bibinfo {volume} {10}},\
  \bibinfo {pages} {5500} (\bibinfo {year} {2019})},\ \bibinfo {note} {number:
  1 Publisher: Nature Publishing Group}\BibitemShut {NoStop}%
\bibitem [{\citenamefont {Shehata}\ \emph {et~al.}(2023)\citenamefont
  {Shehata}, \citenamefont {Simion}, \citenamefont {Li}, \citenamefont
  {Mohiyaddin}, \citenamefont {Wan}, \citenamefont {Mongillo}, \citenamefont
  {Govoreanu}, \citenamefont {Radu}, \citenamefont {De~Greve},\ and\
  \citenamefont {Van~Dorpe}}]{Shehata2023}%
  \BibitemOpen
  \bibfield  {author} {\bibinfo {author} {\bibfnamefont {M.~M. E.~K.}\
  \bibnamefont {Shehata}}, \bibinfo {author} {\bibfnamefont {G.}~\bibnamefont
  {Simion}}, \bibinfo {author} {\bibfnamefont {R.}~\bibnamefont {Li}}, \bibinfo
  {author} {\bibfnamefont {F.~A.}\ \bibnamefont {Mohiyaddin}}, \bibinfo
  {author} {\bibfnamefont {D.}~\bibnamefont {Wan}}, \bibinfo {author}
  {\bibfnamefont {M.}~\bibnamefont {Mongillo}}, \bibinfo {author}
  {\bibfnamefont {B.}~\bibnamefont {Govoreanu}}, \bibinfo {author}
  {\bibfnamefont {I.}~\bibnamefont {Radu}}, \bibinfo {author} {\bibfnamefont
  {K.}~\bibnamefont {De~Greve}},\ and\ \bibinfo {author} {\bibfnamefont
  {P.}~\bibnamefont {Van~Dorpe}},\ }\href
  {https://doi.org/10.1103/PhysRevB.108.045305} {\bibfield  {journal} {\bibinfo
   {journal} {Phys. Rev. B}\ }\textbf {\bibinfo {volume} {108}},\ \bibinfo
  {pages} {045305} (\bibinfo {year} {2023})}\BibitemShut {NoStop}%
\bibitem [{\citenamefont {Choi}\ and\ \citenamefont
  {Joynt}(2023)}]{choi2023interacting}%
  \BibitemOpen
  \bibfield  {author} {\bibinfo {author} {\bibfnamefont {Y.}~\bibnamefont
  {Choi}}\ and\ \bibinfo {author} {\bibfnamefont {R.}~\bibnamefont {Joynt}},\
  }\href@noop {} {\bibinfo {title} {Interacting random-field dipole defect
  model for heating in semiconductor-based qubit devices}} (\bibinfo {year}
  {2023}),\ \Eprint {https://arxiv.org/abs/2308.00711} {arXiv:2308.00711
  [quant-ph]} \BibitemShut {NoStop}%
\bibitem [{\citenamefont {Chan}\ \emph {et~al.}(2018)\citenamefont {Chan},
  \citenamefont {Huang}, \citenamefont {Yang}, \citenamefont {Hwang},
  \citenamefont {Hensen}, \citenamefont {Tanttu}, \citenamefont {Hudson},
  \citenamefont {Itoh}, \citenamefont {Laucht}, \citenamefont {Morello},\ and\
  \citenamefont {Dzurak}}]{chan_assessment_2018}%
  \BibitemOpen
  \bibfield  {author} {\bibinfo {author} {\bibfnamefont {K.~W.}\ \bibnamefont
  {Chan}}, \bibinfo {author} {\bibfnamefont {W.}~\bibnamefont {Huang}},
  \bibinfo {author} {\bibfnamefont {C.~H.}\ \bibnamefont {Yang}}, \bibinfo
  {author} {\bibfnamefont {J.~C.~C.}\ \bibnamefont {Hwang}}, \bibinfo {author}
  {\bibfnamefont {B.}~\bibnamefont {Hensen}}, \bibinfo {author} {\bibfnamefont
  {T.}~\bibnamefont {Tanttu}}, \bibinfo {author} {\bibfnamefont {F.~E.}\
  \bibnamefont {Hudson}}, \bibinfo {author} {\bibfnamefont {K.~M.}\
  \bibnamefont {Itoh}}, \bibinfo {author} {\bibfnamefont {A.}~\bibnamefont
  {Laucht}}, \bibinfo {author} {\bibfnamefont {A.}~\bibnamefont {Morello}},\
  and\ \bibinfo {author} {\bibfnamefont {A.~S.}\ \bibnamefont {Dzurak}},\
  }\href {https://doi.org/10.1103/PhysRevApplied.10.044017} {\bibfield
  {journal} {\bibinfo  {journal} {Physical Review Applied}\ }\textbf {\bibinfo
  {volume} {10}},\ \bibinfo {pages} {044017} (\bibinfo {year} {2018})},\
  \bibinfo {note} {publisher: American Physical Society}\BibitemShut {NoStop}%
\bibitem [{\citenamefont {Unseld}\ \emph {et~al.}(2023)\citenamefont {Unseld},
  \citenamefont {Meyer}, \citenamefont {Mądzik}, \citenamefont {Borsoi},
  \citenamefont {de~Snoo}, \citenamefont {Amitonov}, \citenamefont {Sammak},
  \citenamefont {Scappucci}, \citenamefont {Veldhorst},\ and\ \citenamefont
  {Vandersypen}}]{unseld_2d_2023}%
  \BibitemOpen
  \bibfield  {author} {\bibinfo {author} {\bibfnamefont {F.~K.}\ \bibnamefont
  {Unseld}}, \bibinfo {author} {\bibfnamefont {M.}~\bibnamefont {Meyer}},
  \bibinfo {author} {\bibfnamefont {M.~T.}\ \bibnamefont {Mądzik}}, \bibinfo
  {author} {\bibfnamefont {F.}~\bibnamefont {Borsoi}}, \bibinfo {author}
  {\bibfnamefont {S.~L.}\ \bibnamefont {de~Snoo}}, \bibinfo {author}
  {\bibfnamefont {S.~V.}\ \bibnamefont {Amitonov}}, \bibinfo {author}
  {\bibfnamefont {A.}~\bibnamefont {Sammak}}, \bibinfo {author} {\bibfnamefont
  {G.}~\bibnamefont {Scappucci}}, \bibinfo {author} {\bibfnamefont
  {M.}~\bibnamefont {Veldhorst}},\ and\ \bibinfo {author} {\bibfnamefont
  {L.~M.~K.}\ \bibnamefont {Vandersypen}},\ }\href
  {https://doi.org/10.48550/arXiv.2305.19681} {\bibinfo {title} {A {2D} quantum
  dot array in planar \${\textasciicircum}\{28\}\${Si}/{SiGe}}} (\bibinfo
  {year} {2023}),\ \bibinfo {note} {arXiv:2305.19681 [cond-mat,
  physics:quant-ph]}\BibitemShut {NoStop}%
\bibitem [{\citenamefont {Borsoi}\ \emph {et~al.}(2022)\citenamefont {Borsoi},
  \citenamefont {Hendrickx}, \citenamefont {John}, \citenamefont {Motz},
  \citenamefont {van Riggelen}, \citenamefont {Sammak}, \citenamefont
  {de~Snoo}, \citenamefont {Scappucci},\ and\ \citenamefont
  {Veldhorst}}]{borsoi_shared_2022}%
  \BibitemOpen
  \bibfield  {author} {\bibinfo {author} {\bibfnamefont {F.}~\bibnamefont
  {Borsoi}}, \bibinfo {author} {\bibfnamefont {N.~W.}\ \bibnamefont
  {Hendrickx}}, \bibinfo {author} {\bibfnamefont {V.}~\bibnamefont {John}},
  \bibinfo {author} {\bibfnamefont {S.}~\bibnamefont {Motz}}, \bibinfo {author}
  {\bibfnamefont {F.}~\bibnamefont {van Riggelen}}, \bibinfo {author}
  {\bibfnamefont {A.}~\bibnamefont {Sammak}}, \bibinfo {author} {\bibfnamefont
  {S.~L.}\ \bibnamefont {de~Snoo}}, \bibinfo {author} {\bibfnamefont
  {G.}~\bibnamefont {Scappucci}},\ and\ \bibinfo {author} {\bibfnamefont
  {M.}~\bibnamefont {Veldhorst}},\ }\href
  {https://doi.org/10.48550/arXiv.2209.06609} {\bibinfo {title} {Shared control
  of a 16 semiconductor quantum dot crossbar array}} (\bibinfo {year} {2022}),\
  \bibinfo {note} {arXiv:2209.06609 [cond-mat, physics:quant-ph]}\BibitemShut
  {NoStop}%
\bibitem [{\citenamefont {Gilbert}\ \emph {et~al.}(2023)\citenamefont
  {Gilbert}, \citenamefont {Tanttu}, \citenamefont {Lim}, \citenamefont {Feng},
  \citenamefont {Huang}, \citenamefont {Cifuentes}, \citenamefont {Serrano},
  \citenamefont {Mai}, \citenamefont {Leon}, \citenamefont {Escott},
  \citenamefont {Itoh}, \citenamefont {Abrosimov}, \citenamefont {Pohl},
  \citenamefont {Thewalt}, \citenamefont {Hudson}, \citenamefont {Morello},
  \citenamefont {Laucht}, \citenamefont {Yang}, \citenamefont {Saraiva},\ and\
  \citenamefont {Dzurak}}]{gilbert_-demand_2023}%
  \BibitemOpen
  \bibfield  {author} {\bibinfo {author} {\bibfnamefont {W.}~\bibnamefont
  {Gilbert}}, \bibinfo {author} {\bibfnamefont {T.}~\bibnamefont {Tanttu}},
  \bibinfo {author} {\bibfnamefont {W.~H.}\ \bibnamefont {Lim}}, \bibinfo
  {author} {\bibfnamefont {M.}~\bibnamefont {Feng}}, \bibinfo {author}
  {\bibfnamefont {J.~Y.}\ \bibnamefont {Huang}}, \bibinfo {author}
  {\bibfnamefont {J.~D.}\ \bibnamefont {Cifuentes}}, \bibinfo {author}
  {\bibfnamefont {S.}~\bibnamefont {Serrano}}, \bibinfo {author} {\bibfnamefont
  {P.~Y.}\ \bibnamefont {Mai}}, \bibinfo {author} {\bibfnamefont {R.~C.~C.}\
  \bibnamefont {Leon}}, \bibinfo {author} {\bibfnamefont {C.~C.}\ \bibnamefont
  {Escott}}, \bibinfo {author} {\bibfnamefont {K.~M.}\ \bibnamefont {Itoh}},
  \bibinfo {author} {\bibfnamefont {N.~V.}\ \bibnamefont {Abrosimov}}, \bibinfo
  {author} {\bibfnamefont {H.-J.}\ \bibnamefont {Pohl}}, \bibinfo {author}
  {\bibfnamefont {M.~L.~W.}\ \bibnamefont {Thewalt}}, \bibinfo {author}
  {\bibfnamefont {F.~E.}\ \bibnamefont {Hudson}}, \bibinfo {author}
  {\bibfnamefont {A.}~\bibnamefont {Morello}}, \bibinfo {author} {\bibfnamefont
  {A.}~\bibnamefont {Laucht}}, \bibinfo {author} {\bibfnamefont {C.~H.}\
  \bibnamefont {Yang}}, \bibinfo {author} {\bibfnamefont {A.}~\bibnamefont
  {Saraiva}},\ and\ \bibinfo {author} {\bibfnamefont {A.~S.}\ \bibnamefont
  {Dzurak}},\ }\href {https://doi.org/10.1038/s41565-022-01280-4} {\bibfield
  {journal} {\bibinfo  {journal} {Nature Nanotechnology}\ ,\ \bibinfo {pages}
  {1}} (\bibinfo {year} {2023})},\ \bibinfo {note} {publisher: Nature
  Publishing Group}\BibitemShut {NoStop}%
\bibitem [{\citenamefont {Bourdet}\ and\ \citenamefont
  {Niquet}(2018)}]{Bourdet2018}%
  \BibitemOpen
  \bibfield  {author} {\bibinfo {author} {\bibfnamefont {L.}~\bibnamefont
  {Bourdet}}\ and\ \bibinfo {author} {\bibfnamefont {Y.~M.}\ \bibnamefont
  {Niquet}},\ }\bibfield  {journal} {\bibinfo  {journal} {Physical Review B}\
  }\textbf {\bibinfo {volume} {97}},\ \href
  {https://doi.org/10.1103/PhysRevB.97.155433} {10.1103/PhysRevB.97.155433}
  (\bibinfo {year} {2018}),\ \bibinfo {note} {arXiv: 1802.04693 Publisher:
  American Physical Society}\BibitemShut {NoStop}%
\end{thebibliography}%

\newpage



\begin{table*}[]

\centering
\begin{tabular}{cccccccc}
\multicolumn{1}{l}{\textbf{Device}} & \textbf{Type of Device} & \multicolumn{1}{c}{\textbf{\begin{tabular}[c]{@{}c@{}}Electron\\ numbers\end{tabular}}} & \multicolumn{1}{c}{\textbf{\begin{tabular}[c]{@{}c@{}}Qubits\\ below\end{tabular}}} & \multicolumn{1}{c}{\textbf{$\mu_B dg_1/dV_{P1}$}} & \multicolumn{1}{c}{\textbf{$\mu_B dg_2/dV_{P2}$}} & \multicolumn{1}{c}{\textbf{$\mu_B dg_2/dV_{P1}$}} & \multicolumn{1}{c}{\textbf{$\mu_B dg_1/dV_{P2}$}} \\
\hline
\textbf{A} & Three Dot Device & (3,1,0) & P1 , P2 & 25 & -50 & -22.5 & 25 \\
\textbf{B} & Three   Dot Device & (1,1,0) & P1 , P2 & 17 & 5.6 & -1.7 & -0.6 \\
\textbf{B} & Three Dot Device & (1,3,0) & P1 , P2 & -11.7 & -45 & 4.5 & 0.9 \\
\textbf{B} & Three   Dot Device & (3,1,0) & P1 , P2 & 9.1 & 2.3 & -0.1 & -3.2 \\
\textbf{B} & Three Dot Device & (3,3,0) & P1 , P2 & -2.7 & -40.5 & 0.8 & 4.2 \\
\textbf{B} & Three   Dot Device & (3,5,0) & P1 , P2 & 20.1 & -14 & 3.4 & -5.5 \\
\textbf{C} & Four Dot Device & (0,0,3,5) & P3 , P4 & 0.9 & -3.2 & -3.2 & 0.6 \\
\textbf{D} & Three   Dot Device & (0,3,1) & P2 , P3 & 10.8 & 23.0 & -5.0 & -3.2 \\
\textbf{E} & Three Dot Device & (0,3,1) & P2 , P3 & -10.4 & 49 & -4.2 & 0 \\
\textbf{F} & Three   Dot Device & (3,1,0) & P2 , P3 & 20.6 & -21.8 & 14.5 & 9.3 \\
\textbf{G} & Four Dot Device & (0,1,1,3) & P3 , P4 & 6.2 & -11.3 & -2.6 & -10.5 \\
\textbf{H} & Four Dot Device & (0,1,3,0) & P2 , P3 & -2.1 & -24.3 & -0.9 & 10.9 \\
\textbf{I} & Three Dot Device & (3,1,0) & P1 , P2 & -24.2 & 12.4 & 3.3 & 0 \\
\textbf{I} & Three   Dot Device & (1,3 ,0) & P1 , P2 & -1 & -11 & 8.8 & 19.3 \\
\textbf{I}	& Three Dot Device	& (1,1,0) & P1 , P2 &	-7.8 &		14 & 9.2 &	-5.6 

\end{tabular}
\caption{$\vert$ \label{tab:StarkShifts} Table of Stark shifts measured in 9 different devices. The units for Stark shift data are $[\mu_B dg_2/dV_{P1}] =$~[MHz~V$^{-1}$~T$^{-1}$]. In some devices we measured spin qubits at different electron configurations. This table includes data from spin qubits formed at the outer-shell of quantum dots with 1, 3 and 5 electrons\cite{Leon2020}. We also include data from two different generations of devices, one with three plunger gates to form quantum dots and the other with 4 plunger gates (P1, P2 , P3 , P4)~\cite{tanttu_stability_2023}. The fourth column indicates the gate below which the two spin qubits are formed. Qubits formed at different electronic configurations are assumed to be different under this table, even if they are formed in the same device and under the same gate. This is because changes in the potential profile and spin density can affect the surface profile of the quantum dots. The total qubit count is 30 under these assumptions. }

\end{table*}

\begin{table*}[h]
\centering
\begin{tabular}{ccccc}
 & \textbf{$dx_1/dV $}    & \textbf{$dE_{z1}/dV$}    & \textbf{$dx_2/dV$}    & \textbf{$dE_{z2}/dV$}   \\
 \textbf{Gates} & [nm V$^{-1}$] & [eV nm$^{-1}$V$^{-1}$] & [nm V$^{-1}$] & [eV nm$^{-1}$V$^{-1}$]\\
\hline
P1   & {-6.74} & {13.42} & {-2.95} & { -2.11}   \\
P2   & {4.95} & { -0.68}   & { 5.5} & { 14.75} \\
J1   & { 6.88} & { 0.46}   & { -3.57} & { -0.22}   \\
P3   & { 0.04}   & { -0.02}   & {0.13} & { 0.06} \\
J2   & { 0.02}   & { -0.01}   & {0.13} & { 0.06}   \\
UB   & { 5.88 y-axis}   & { -4.5}   & { 6.37 y-axis} & { -4.96} 

\end{tabular}

\caption{$\vert$ \label{tab:PotentialSweeps} Electrostatic simulations in Comsol Multiphysics: Summary of the action of each gate action on the quantum dots obtained from fits of electrostatic simulations of realistic devices to the harmonic model \eqref{eq2}. Details can be found in~\cite{cifuentes_bounds_2023}. }
\end{table*}

\end{document}